\documentclass[fleqn,usenatbib]{mnras}

\usepackage{newtxtext,newtxmath}
\usepackage[T1]{fontenc}

\DeclareRobustCommand{\VAN}[3]{#2}
\let\VANthebibliography\thebibliography
\def\thebibliography{\DeclareRobustCommand{\VAN}[3]{##3}\VANthebibliography}

\usepackage{graphicx}	
\usepackage{amsmath}	
\usepackage{natbib}
\usepackage{hyperref}
\usepackage[table,xcdraw]{xcolor} 
\usepackage{longtable}
\usepackage{array}
\usepackage{tabularx}
\usepackage{longtable}
\usepackage{booktabs}
\usepackage{graphicx}
\usepackage{geometry}
\usepackage{colortbl}
\usepackage{xcolor}
\usepackage{makecell}
\usepackage{booktabs}
\usepackage{multicol}
\usepackage{svg}
\usepackage[table,xcdraw]{xcolor}
\newcommand{\xcell}{\cellcolor{lightgray}}
\DeclareUnicodeCharacter{9EC4}{Huang}
\DeclareUnicodeCharacter{7855}{Shuo}

\title[M.L. for Biosignature Classification]{Machine-assisted classification of potential biosignatures in Earth-like exoplanets using low signal-to-noise ratio transmission spectra}

\author[David S. Duque-Castaño et al.]{
David S. Duque-Castaño,$^{1}$\thanks{E-mail: \href{mailto:dsantiago.duque@udea.edu.co}{dsantiago.duque@udea.edu.co}, \href{mailto:dasan.academico@gmail.com}{dasan.academico@gmail.com} \newline \href{http://orcid.org/0000-0003-3614-7904}{ORCID: 0000-0003-3614-7904}}
Jorge I. Zuluaga,$^{1}$\thanks{\href{http://orcid.org/0000-0002-6140-3116}{ORCID: 0000-0002-6140-3116}}
Lauren Flor-Torres$^{1}$\thanks{\href{http://orcid.org/0000-0003-4134-9615}{ORCID: 0000-0003-4134-9615}}
\\
$^{1}$SEAP/FACom, Instituto de F\'{\i}sica - FCEN, Universidad de Antioquia, Calle 70 No. 52-21, Medell\'in, Colombia\\
}

\date{Accepted XXX. Received YYY; in original form ZZZ}

\pubyear{\the\year{}}

\begin{document}
\label{firstpage}
\pagerange{\pageref{firstpage}--\pageref{lastpage}}
\maketitle

\begin{abstract}
The search for atmospheric biosignatures in Earth-like exoplanets is one of the most pressing challenges in observational astrobiology. Detecting biogenic gases in terrestrial planets requires high-resolution observations and long integration times.
In this work, we developed and tested a general machine-learning methodology designed to classify transmission spectra with low Signal-to-Noise Ratio (SNR) according to their potential to contain biosignatures or bioindicators.
To achieve this, we trained a set of models capable of classifying noisy transmission spectra (including stellar contamination) as containing methane, ozone, and/or water (multilabel classification), or simply as being interesting for follow-up observations (binary classification). The models were trained using $\sim10^7$ synthetic spectra of planets similar to TRAPPIST-1 e, generated with the package \href{https://pypi.org/project/multirex/}{\tt MultiREx}.
The trained algorithms correctly classified most of the test planets with transmission spectra having an SNR as low as 4, containing methane and/or ozone at mixing ratios similar to those of modern and Proterozoic Earth. Tests on realistic synthetic spectra, based on the current Earth's atmosphere, indicate that some of our models would classify most inhabited terrestrial planets observed with JWST/NIRSpec PRISM around M-dwarfs at distances similar to or smaller than that of TRAPPIST-1 e as likely to contain bioindicators, using 4–10 transits.
These results have significant implications for the design of observing programs and future campaigns. Machine-assisted strategies, such as the one presented here, could greatly optimize the use of JWST resources for biosignature and bioindicator searches, while maximizing the chances of a real discovery through dedicated follow-up observations of promising candidates.
\end{abstract}




\section{Introduction}\label{introducciuxf3n}

One of the strategies conceived to search for life beyond Earth involves detecting signals on exoplanets that may be closely associated with life and have a low probability of being abiotic in origin. We call these signals, biosignatures (see e.g. \citealt {schwietermanExoplanetBiosignaturesReview2018,desmaraisNASAAstrobiologyRoadmap2008} and references therein). Currently, our instrumental capabilities allow us to detect biosignatures that have a global impact, especially those concentrated in the atmosphere and observable through spectroscopy (see e.g. \citealt{schwietermanExoplanetBiosignaturesReview2018}). Furthermore, in cases where the likelihood of an abiotic origin is not negligible, the concept of a bioindicator can be utilised to distinguish molecular species that are more likely to be produced through biological processes. Bioindicators, while not definitive evidence of life on their own, provide valuable complementary clues that help to discern biological activity from abiotic processes. 

However, detecting biosignatures on rocky exoplanets within the habitable zone of their stars poses significant challenges due to the low signal-to-noise ratio resulting from the smaller relative radii of the star-to-planet. M-dwarfs offer an opportunity to study the atmospheres of Earth-like planets. Although these stars may exhibit high X-ray and ultraviolet (XUV) activity in their early stages, secondary atmospheres may remain stable as they age \citep{franceHighenergyRadiationEnvironment2020}. It is well established that M-dwarfs exhibit greater transit depths compared to other stellar types \citep{wunderlichDetectabilityAtmosphericFeatures2019} increasing the chances of detecting and analysing exoplanetary atmospheres. Instruments such as the James Webb Space Telescope (JWST) demonstrate our ability to reliably detect atmospheric species on exoplanets around M-dwarfs, such as possible H$_2$-rich atmospheres \citep{madhusudhanCarbonbearingMoleculesPossible2023, bennekeJWSTRevealsCH$_4$2024}, and even showcase the capability to identify, as the most likely scenario, Earth-like atmospheres such as N$_2$- or CO$_2$-dominated atmospheres \citep{cadieuxTransmissionSpectroscopyHabitable2024}.

Recent simulations on the detectability of an Earth-like atmosphere using the JWST \citep{barstowHabitableWorldsJWST2016, wunderlichDetectabilityAtmosphericFeatures2019, linDifferentiatingModernPrebiotic2021, lustig-yaegerEarthTransitingExoplanet2023} have revealed a challenging scenario. To detect robust bioindicators such as ozone (O$_3$) (for the distinction between bioindicators and biosignatures, see \autoref{sec:biosignatures}), a large number of transits is required (e.g., up to 200 transits in the case of TRAPPIST-1 e; see \citealt{linDifferentiatingModernPrebiotic2021}) to achieve statistically significant detections. Despite this challenge, detecting methane (CH$_4$) and water vapour (H$_2$O) presents a promising opportunity. Studies have demonstrated that, using a reasonable number of transits, the presence of these atmospheric species, which are typically associated with a global biosphere, can be retrieved \citep{wunderlichDetectabilityAtmosphericFeatures2019, linDifferentiatingModernPrebiotic2021, lustig-yaegerEarthTransitingExoplanet2023}. However, it is important to note that CH$_4$ is a less robust bioindicator compared to O$_3$ (for more details on the robustness of ozone as a bioindicator, see \autoref{sec:bio-O}).

Considering the substantial resource requirements associated with conducting such observational campaigns, a more effective strategy may be to allocate JWST time to conduct a low signal-to-noise ratio (SNR) survey. Although this may not allow for statistically significant retrievals, it would facilitate planning future follow-up observations of promising candidate targets with current and future, more powerful telescopes (e.g., ELT, LUVOIR, HabEx, Roman, ARIEL).

In this paper, we propose and numerically test methods and tools to support such a strategy. For this purpose, we design and test a set of machine learning (ML) methods and tools aimed at labelling low SNR spectra that may contain interesting biosignatures or bioindicators. It is important to stress that the tools presented here are not designed to perform a retrieval of the abundances of chemical species, but to identify interesting candidates for follow-up observations.

The use of ML in analysing exoplanet spectra has garnered significant attention in recent literature. Multiple algorithms have been developed and trained on diverse datasets and objectives \citep{ marquez-neilaSupervisedMachineLearning2018,soboczenskiBayesianDeepLearning2018,zingalesExoGANRetrievingExoplanetary2018, cobbEnsembleBayesianNeural2019,nixonAssessmentSupervisedMachine2020}. Primarily, these algorithms have been employed as alternatives to the computationally intensive Bayesian retrieval methods \citep{munsaketRetrievingExoplanetAtmospheric2021, ardevolmartinezConvolutionalNeuralNetworks2022,vasistNeuralPosteriorEstimation2023,ardevolmartinezFlopPITyEnablingSelfconsistent2024}. In \autoref{tbl:mlearning_models} we summarise some of the strategies that have been devised for performing atmospheric composition retrieval and are available in the literature.


\newcommand{\impwidth}{0.18\textwidth}
\newcommand{\molwidth}{0.0238\textwidth}

\begin{table*}
\caption{Studies on the use of ML algorithms in exoplanet spectra. We describe the purpose and application of the algorithms used in each study. Additionally, we list the exoplanet regimes each algorithm focuses on and indicate whether they were trained or tested with C/O ratios plus metallicity (represented in the column C/O), or the most common free chemistry molecules.}
\label{tbl:mlearning_models}
\centering
\begin{tabularx}{\textwidth}{
  |
  >{\centering\arraybackslash\scriptsize}p{\impwidth}  
  >{\centering\arraybackslash\scriptsize}p{\impwidth}  
  >{\centering\arraybackslash\scriptsize}p{\impwidth}|  
  >{\centering\arraybackslash\scriptsize}p{\molwidth}|  
  >{\centering\arraybackslash\scriptsize}p{\molwidth}|  
  >{\centering\arraybackslash\scriptsize}p{\molwidth}|  
  >{\centering\arraybackslash\scriptsize}p{\molwidth}|  
  >{\centering\arraybackslash\scriptsize}p{\molwidth}|  
  >{\centering\arraybackslash\scriptsize}p{\molwidth}|  
  >{\centering\arraybackslash\scriptsize}p{\molwidth}|  
  >{\centering\arraybackslash\scriptsize}p{\molwidth}|  
}
\hline
\small\textbf{Reference} & \small\textbf{ML Method(s)} & \small\textbf{Targets} & \textbf{C/O} & \textbf{H$_2$O} & \textbf{CO$_2$} & \textbf{CO} & \textbf{CH$_4$} & \textbf{NH$_3$} & \textbf{HCN} & \textbf{O$_3$} \\
\hline
\citet{gebhardParameterizingPressureTemperature2024} & Convolutional Neural Networks and Multilayer Perceptrons (Regression) & Hot Jupiters and Earth-like Planets & & & & & & & & \\
\hline
\citet{ardevolmartinezFlopPITyEnablingSelfconsistent2024} & Sequential Neural Posterior Estimation with Normalizing Flows (Regression) & Wide range of planets, Brown Dwarfs & \xcell & \xcell & \xcell & & & & & \\
\hline
\citet{forestanoSearchingNovelChemistry2023} & Local Outlier Factor, One-Class Support Vector Machine (Unsupervised Machine Learning, Outlier Detection) & Hot Jupiters & & \xcell & \xcell & \xcell & \xcell & \xcell & & \\
\hline
\citet{vasistNeuralPosteriorEstimation2023} & Neural Posterior Estimation with Normalizing Flows (Regression) & Gas Giants & \xcell & & & & & & & \\
\hline
\citet{ardevolmartinezConvolutionalNeuralNetworks2022} & Convolutional Neural Networks (Regression) & Gas Giants & \xcell & \xcell & \xcell & \xcell & \xcell & \xcell & & \\
\hline
\citet{himesAccurateMachinelearningAtmospheric2022} & Convolutional Neural Networks (Regression) & Hot Jupiters & & \xcell & \xcell & \xcell & \xcell & & & \\
\hline
\citet{matchevUnsupervisedMachineLearning2022} & k-means, PCA and ISOMAP (Unsupervised) & Hot Jupiters & & \xcell & & & & \xcell & \xcell & \\
\hline
\citet{matchevAnalyticalModelingExoplanet2022} & Symbolic Regression & Hot Jupiters & & & & & & & & \\
\hline
\citet{munsaketRetrievingExoplanetAtmospheric2021} & Random Forest (Supervised ML, Regression) & Hot Jupiters & \xcell & & & & & & & \\
\hline
\citet{yipPeekingBlackBox2021} & Multilayer Perceptrons, Convolutional Neural Networks and Long Short-Term Memory Networks (Regression) & Wide range of planets & & \xcell & \xcell & \xcell & \xcell & \xcell & & \\
\hline
\citet{guzman-mesaInformationContentJWST2020} & Random Forest (Regression) & Warm Neptunes & & \xcell & \xcell & \xcell & \xcell & \xcell & \xcell & \\
\hline
\citet{hayesOptimizingExoplanetAtmosphere2020} & K-means, Principal Component Analysis (Unsupervised Classification) & Jupiter-like planets & \xcell & & & & & & & \\
\hline
\citet{nixonAssessmentSupervisedMachine2020} & Random Forest (Regression) & Hot Jupiters & & \xcell & & & & \xcell & \xcell & \\
\hline
\citet{cobbEnsembleBayesianNeural2019} & Bayesian Neural Network (Regression) & Hot Jupiters & & \xcell & & & & \xcell & \xcell & \\
\hline
\citet{soboczenskiBayesianDeepLearning2018} & Convolutional Neural Networks with Monte Carlo Dropout (Regression) & Rocky terrestrial exoplanets & & \xcell & \xcell & \xcell & \xcell & \xcell & & \xcell \\
\hline
\citet{marquez-neilaSupervisedMachineLearning2018} & Random Forest (Regression) & Hot Jupiters & & \xcell & & \xcell & \xcell & \xcell & \xcell & \\
\hline
\citet{zingalesExoGANRetrievingExoplanetary2018} & Generative Adversarial Networks (Regression) & Hot Jupiters & \xcell & \xcell & \xcell & \xcell & \xcell & & & \\
\hline
\citet{waldmannDREAMINGATMOSPHERES2016} & Deep-Belief Networks (Classification) & Wide range of planets & & \xcell & \xcell & \xcell & \xcell & \xcell & \xcell & \\
\hline
\end{tabularx}
\end{table*}

Additionally, the capacity of ML to support parts of the retrieval process – such as suggesting priors \citep{hayesOptimizingExoplanetAtmosphere2020}, identifying molecules \citep{waldmannDREAMINGATMOSPHERES2016}, using neural network parametrisation of pressure-temperature profiles for better efficiency and physical consistency \citep{gebhardParameterizingPressureTemperature2024}, or even replacing the spectrum generation process through radiative transfer – has been explored \citep{himesAccurateMachinelearningAtmospheric2022}. Conversely, unsupervised learning techniques have also been proposed for analysing spectroscopic data and detecting anomalous chemical compositions \citep{forestanoSearchingNovelChemistry2023, matchevUnsupervisedMachineLearning2022}. ML has been used to assess molecule detection capabilities in various instruments of the JWST, demonstrating how machine learning techniques have become robust alternatives to classical retrieval methods \citep{guzman-mesaInformationContentJWST2020}.

Despite these advancements, traditional ML methods are often seen as black boxes, where the internal workings and the features driving the models are not easily interpretable, limiting the ability to achieve physical understanding. Efforts have been made to address this issue. For instance, \citet{yipPeekingBlackBox2021} developed tools to elucidate the functioning of ML methods when making predictions. Similarly, \citet{matchevAnalyticalModelingExoplanet2022} demonstrated the use of symbolic regression for characterising transit spectra, enabling a physical understanding of the problem.

Furthermore, these studies have primarily focused on gaseous planets, especially Hot Jupiters like WASP-12. Only a minority of studies focus on rocky and Earth-like planets, indicating a significant gap in the training data available for these types of planets. Nonetheless, recent research has started to consider brown dwarfs, as seen in the work of \citet{ardevolmartinezFlopPITyEnablingSelfconsistent2024} and \citet{lueberIntercomparisonBrownDwarf2023}. The implemented regression techniques commonly include Random Forest and various types of neural networks. Additionally, unsupervised learning techniques are particularly focused on clustering and anomaly detection.

Based on our literature review, to date, only one study has been conducted specifically on classifying potential molecules in spectra \citep{waldmannDREAMINGATMOSPHERES2016}. The training datasets contain atmospheric compositions defined by C/O ratio and metallicity, or free chemistry, which includes molecules such as H$_2$O, NH$_3$, CH$_4$, CO$_2$, and CO (see \autoref{tbl:mlearning_models}).

As pointed out before, our proposal is not to perform the full retrieval, which—even using ML techniques—requires high SNR signals. This is demonstrated by the fact that many of the previously cited works are related to giant planets. However, if we can obtain low SNR spectra and identify intriguing Earth-like planets, we can concentrate our search on the most promising candidates.

This paper is organised as follows. In \autoref{sec:SML} we describe the ML methods used in this paper, particularly introducing the language and quantities used in classification-supervised machine learning. To test our methods, we focus on a very interesting target, namely TRAPPIST-1 e, which is described in \autoref{sec:trappist1e}. \autoref{sec:biosignatures} is dedicated to describing the most interesting species on which our numerical experiments are focused. Furthermore, \autoref{sec:stellar-auto} provide a comprehensive discussion of stellar contamination—an effect that must be taken into account when assessing how it affects the observed transit spectrum by introducing spurious features that might be mistakenly attributed to the planet—and outline the autoencoder approach we use to address it. The methods, tools, and preliminary results are detailed in \autoref{sec:gen-data} and \autoref{sec:RF}. Contextualisation of the results based on the required transits using the NIRSpec MIRI instrument of JWST is discussed in \autoref{sec:SNRtransits}. Our algorithms are tested using realistic spectra of Earth in \autoref{sec:realistic}. The limitations of our approach and the training data are discussed in \autoref{sec:discussion}. Finally, the conclusions of our numerical experiments are drawn in \autoref{sec:conclusion}. All the tools and data developed for this paper are publicly available, with access information provided in \autoref{sec:data-availability}.


\section{Supervised Machine Learning for biosignatures detection}
\label{sec:SML}

Machine learning for classification (MLC) is an approach to building a predictive model that assigns a label or category to a data point or instance based on a set of known features (see, e.g., \citealt{geronHandsonMachineLearning2023}). For example, we aim to label planets based on the presence of biosignatures/bioindicators using simple categories: O$_3/\text{\tt No-}$O$_3$, CH$_4$/$\text{\tt No-}$CH$_4$.

In MLC, a training set consists of labelled observations (e.g., a list of planets with and without biosignatures/bioindicators), where each observation comprises features and a classification label. The simplest form is binary classification, where an instance either belongs to a category or does not. The goal is for the classifier model to accurately predict the classification label for new, unlabelled data, enabling decisions based on the classification of such data. Our objective is for the trained algorithm to identify which planets have a specific biosignature and which do not, using a low SNR spectrum.


\subsection{Random forest for biosignature classification}

Many algorithms have been devised for MLC. Here, we focus on Random Forest (RF), a family of methods that is not only simple enough but also particularly well-suited to our requirements. RF demonstrates robustness to noise by maintaining high accuracy through the aggregation of predictions from multiple trees, thereby reducing the impact of noisy instances. Additionally, RF is inherently resistant to overfitting due to the Law of Large Numbers, which ensures the convergence of the model’s error as more trees are added \citep{breimanRandomForests2001}. This characteristic makes RF especially effective for complex datasets with numerous variables, each contributing marginally to the prediction, thus providing a robust solution for high-dimensional data classification.

In our case, RF takes signals from various spectral bands (these are the features) and creates decision trees based on the values of these specific signals. Each decision tree in the forest can use multiple features simultaneously to make decisions, determining whether these signals contribute to identifying a biosignature.


\subsection{Metrics for classification}
\label{sec:metrics}

In classification problems, the performance of a machine learning algorithm is not only measured in terms of how many planets were correctly classified (``True'' cases), i.e. the accuracy, or how many were not (``False'' cases), i.e. the errors.  It is also necessary to categorise the classification results into positives and negatives: either True Positives (TP) or True Negatives (TN), which are related to accuracy; and into False Positives (FP) or False Negatives (FN), which correspond to errors. This categorization is summarised in what is called a {\it confusion matrix} \citep{kelleherFundamentalsMachineLearning2015}. 

\begin{figure}
\centering
\includegraphics[scale=0.43]{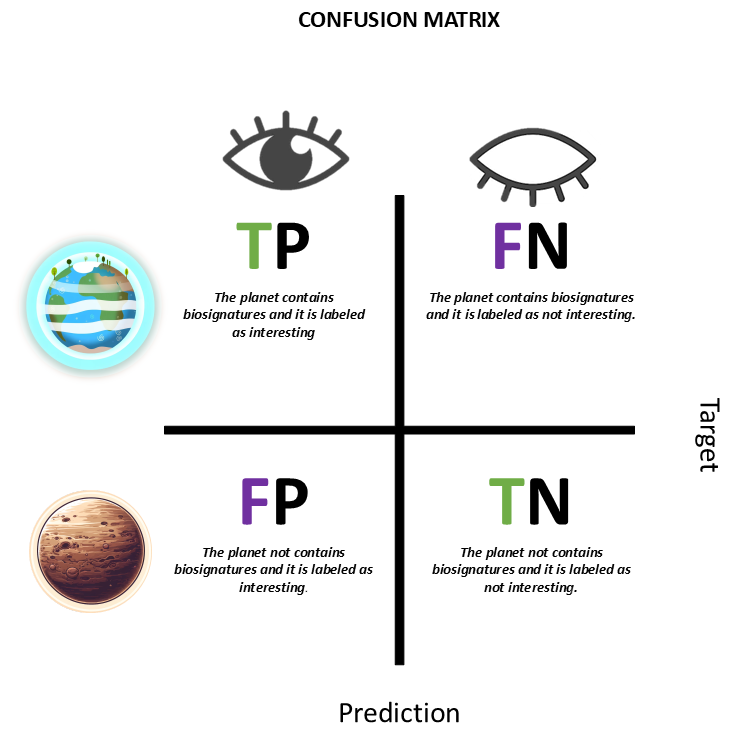}
\caption{Schematic representation of the confusion matrix employed for ML classification in the context of biosignature searching. Instances (planets) in the first row possess biosignatures/bioindicators, while those in the second row do not. An algorithm labels instances (planets) in the first column as potentially having biosignatures (i.e., interesting), whereas those in the second column are not considered interesting. This framework illustrates the different classification outcomes, including true positives (TP), false positives (FP), true negatives (TN), and false negatives (FN), in the search for biosignatures/bioindicators.}\label{fig:mconfusion}
\end{figure}

In \autoref{fig:mconfusion} we schematically represent the categorization in a confusion matrix and, more importantly, illustrate what each case means in the context of potentially biosignature-bearing exoplanets. In the diagram, we introduce the category \textit{interesting} to distinguish planets that deserve follow-up observations or in-depth analysis.We should recall that this is the focus of this work: we do not aim to detect biosignatures/bioindicators using ML but rather to label planets as interesting or not.

Different types of errors carry specific implications for biosignature searching. For example, mistakenly classifying a planet that contains biosignatures/bioindicators as not interesting (FN), could mean the loss of valuable research opportunities. In contrast, erroneously considering a planet without biosignatures/bioindicators as interesting (FP) could lead to inefficient resource allocation. Therefore, understanding the difference between these types of errors is essential for evaluating the accuracy of a model and for adjusting decision and exploration strategies based on the model's classification outcomes (see \autoref{sec:RF}).

Given the limitations of using accuracy alone, i.e. number of correct predictions divided by the total number of predictions, especially when different errors have disproportionately large impacts, more nuanced metrics become indispensable \citep{kelleherFundamentalsMachineLearning2015}. For instance, \emph{recall} which is defined as 

\begin{equation}
    \text{Recall} = \frac{TP}{TP + FN} 
    \label{eq:recall}
\end{equation}

measures the model's efficiency in detecting all relevant instances.  Ideally, a $\text{Recall} = 1$ means that the algorithm correctly classifies all planets that have biosignatures/bioindicators (first row in \autoref{fig:mconfusion}) as interesting exoplanets. In this case, the follow-up observations will be maximally successful. For this reason, we call Recall a \textit{wasting metric}: when Recall decreases, we are excluding promising planets from future observational campaigns and wasting research opportunities.

On the other hand, we have the metric \emph{Precision}, which is defined as

\begin{equation} \text{Precision} = \frac{TP}{TP + FP} \label{precision} \end{equation}

that assesses the accuracy of interesting predictions. This metric is more difficult to interpret since it focuses on the classification rather than the actual presence of biosignatures/bioindicators. When the number of planets without biosignatures/bioindicators classified as interesting (FP) decreases, Precision increases. Therefore, we call Precision a \textit{time-saving metric}, as it helps minimise the time spent studying planets without biosignatures/bioindicators.

In summary, we expect that an effective algorithm applied to a given dataset would have high values for both metrics, i.e., it will be minimally wasteful and maximally time-saving.

But there is a third metric, one that quantifies another aspect of the problem: the discovery opportunity. For this purpose, we use the \emph{F1 Score}, defined as:

\begin{equation} \text{F1 Score} = 2 \cdot \frac{\text{Precision} \times \text{Recall}}{\text{Precision} + \text{Recall}} \label{f1} \end{equation}

In the worst-case scenario, $\text{F1 Score}=0$, meaning that the algorithm is utterly incapable of classifying as interesting those planets that actually have biosignatures/bioindicators. On the other extreme, the ideal case, $\text{F1 Score}=1$, occurs when the algorithm achieves $\text{Recall}=\text{Precision}=1$. From this ideal case, the $\text{F1 Score}$ decreases as the number of FN increases; that is, this metric is very sensitive to instances where a planet with a biosignature is classified as not interesting, thereby losing the discovery opportunity. This is why we call the $\text{F1 Score}$ a \textit{discovery metric}.

All of the previous metrics are focused on positive classifications. However, for the specific purpose of searching for biosignatures/bioindicators, and especially for studying the potential confusion that can arise when an algorithm is designed to detect a specific biosignature in the presence of other molecules, we should use metrics focused on the negative cases. Thus, for example, the \emph{True Negative Rate} (TNR):

\begin{equation} \text{TNR} = \frac{TN}{FP + TN} \label{tnr} \end{equation}

We call this metric a \textit{confusion metric}, in the sense that when the TNR decreases, the algorithm labels planets without biosignatures/bioindicators as interesting. This confusion can arise, for instance, when a biologically irrelevant molecule is “identified” by the algorithm as a biosignature. Essentially, TNR is similar to recall but focuses on the negative labels, ensuring that planets without biosignatures/bioindicators are correctly identified as not interesting. Therefore, TNR can also be understood as a \textit{negative recall}.

Since understanding the definition, meaning and application of a given metric can be very confusing, we have summarised them in \autoref{tab:metrics} with special attention on highlighting their role in the context of biosignature search.

\begin{table*}
\renewcommand{\arraystretch}{1.2} 
\begin{center}
\caption{Summary of metrics used in the evaluation of biosignature detection algorithms. The table includes the metric name, a nickname describing its utility, its specific use in the context of evaluating planets for biosignatures/bioindicators, and extreme cases illustrating the possible values and interpretations of each metric.\label{tab:metrics}}
\begin{tabularx}{\textwidth}{|@{}p{2cm}@{\hspace{2mm}}|@{\hspace{2mm}}p{2cm}@{\hspace{2mm}}|@{\hspace{2mm}}p{4.5cm}@{\hspace{2mm}}|@{\hspace{2mm}}p{8.1cm}@{\hspace{2mm}}|}
\hline\hline
\textbf{Metric} & \textbf{Nickname} & \textbf{Utility} & \textbf{Extreme cases} \\
\hline\hline
Recall & Wasting metric & When small we are excluding promising planets from future observational campaigns, and wasting research opportunities & \begin{itemize} 
    \item Recall = 1: all planets with biosignatures/bioindicators are labelled as interesting 
    \item Recall = 0: no planets with biosignatures/bioindicators are labelled as interesting 
\end{itemize} \\
\hline
Precision & Time-saving metric & Helps to minimize the time spent studying planets without biosignatures/bioindicators & \begin{itemize} 
    \item Precision = 1: All planets labelled as interesting truly have biosignatures/bioindicators.
    \item Precision = 0: Any planet is labelled as interesting even without biosignatures/bioindicators.
\end{itemize} 
\textbf{Note:} Even if the false positives are at their maximum, meaning all non-interesting planets are incorrectly labelled as interesting, the precision is not zero as long as there are true positives. This occurs because precision is calculated as the ratio of true positives to the sum of true positives and false positives. Thus, the presence of any correctly identified interesting planets ensures that precision remains above zero. \\
\hline
F1 Score & Discovery metric & Quantifies the discovery opportunity by balancing Precision and Recall. MAXIMIZE the opportunity to achieve a successful discovery & \begin{itemize} 
    \item F1 Score = 1: when Recall = Precision = 1, meaning the algorithm has perfect precision and \text{Recall}, successfully identifying all interesting planets without any false positives 
    \item F1 Score = 0: algorithm cannot classify any interesting planets correctly 
\end{itemize} Note: When TNR = 0, F1 Score is not necessarily very low if TP is high. \\
\hline
TNR & Confusion metric & Helps to ensure that planets without biosignatures/bioindicators are not mislabelled as interesting, thereby avoiding wasted resources on unpromising candidates & \begin{itemize} 
    \item TNR = 1: all non-interesting planets are correctly identified as not interesting 
    \item TNR = 0: all non-interesting planets are incorrectly labelled as interesting 
\end{itemize} \\
\hline
Hamming Loss & Error rate metric & It measures the fraction of incorrectly predicted molecules, indicating the algorithm's deficiency in misclassifying or omitting the presence of molecules in general.& \begin{itemize} 
    \item Hamming Loss = 0: All predicted labels (biosignatures/bioindicators) are correct.
    \item Hamming Loss = 1: All predicted labels (biosignatures/bioindicators) are incorrect.
\end{itemize} \\
\hline
Exact Match Ratio & Perfect match metric & Measures how well the algorithm classifies all biosignatures/bioindicators for planets correctly at the same time. This metric is very demanding as it requires perfect classification for all labels. Higher values indicate better performance. & \begin{itemize} 
    \item Exact Match Ratio = 1: The algorithm correctly assigns all biosignatures/bioindicators to all planets.
    \item Exact Match Ratio = 0: The algorithm fails to assign the correct set of biosignatures/bioindicators to any planet.
\end{itemize} \\
\hline
\end{tabularx}
\end{center}
\end{table*}


\subsection{Multilabel classification}
\label{subsec:multilabel}

Multilabel classification expands the challenges of traditional classification by allowing each instance, i.e., each planet in the sample, to be linked with multiple labels simultaneously, e.g., potentially having ozone, potentially having CO$_2$, etc. This increases complexity by requiring models to manage interdependencies among multiple labels. 

According to \cite{sorowerLiteratureSurveyAlgorithms2010}, conventional methods are adapted for these scenarios by transforming the problem into multiple binary classification problems or by adapting existing algorithms to process multiple labels. Among the metrics for evaluating these models is Hamming Loss,

\begin{equation}
\text{Hamming Loss} = \frac{1}{N \times L} \sum_{i=1}^{N} \sum_{j=1}^{L} I(y_{ij} \neq \hat{y}_{ij}), \label{eq:hammingloss}
\end{equation}
where \(N\) is the total number of instances (i.e., planets), \(L\) is the number of labels per instance (i.e., the number of molecular species), and \(I\) is the indicator function that returns 1 if the condition is true and 0 otherwise. The condition for this metric is that the $j$th label the algorithm associates with the $i$th instance, \(\hat{y}_{ij}\), is not the correct one, \(y_{ij}\). For instance, if the algorithm predicts that planet 1 has molecule 2 when it does not, then $I(y_{12}\neq\hat{y}_{12})=1$.

Hamming Loss measures the fraction of incorrectly predicted labels. For example, if we need to determine whether a planet has three molecular species and the algorithm incorrectly assigns two of the three labels, the Hamming Loss is $0.\overline{6}$. A large Hamming Loss generally indicates worse multilabel classification performance.

On the other hand, we can use Exact Match Ratio,

\begin{equation}
\text{Exact Match Ratio} = \frac{1}{N} \sum_{i=1}^{N} I(Y_i = \hat{Y}_i), \label{eq:exactmatch}
\end{equation}
where the condition $\hat{Y}_i=Y_i$ implies that the algorithm correctly assigns all possible labels to the $i$th instance.

It is worth mentioning that, besides evaluating Hamming Loss and Exact Match Ratio in a multilabel context, we can combine the metrics defined in previous subsections to compute averages per label, known as \textit{macro averaging}. For instance, if we are classifying multiple molecular species in a sample of planets, we can independently calculate the F1 Score for each molecule and then average them.

\bigskip

In summary, evaluating the performance of a classification algorithm is not a trivial matter. Depending on which aspects of the classification we want to focus on, different metrics can be used. In this work, we will apply a combination of binary and multilabel classification metrics to assess the capabilities of our algorithms.

\section{The case of the star TRAPPIST-1 and its planets}
\label{sec:trappist1e}

In order to test our methods and techniques, we need to select an appropriate investigation case that lies between an idealised, illustrative planet and a genuine target. After considering several possibilities, the TRAPPIST-1e system was identified as an ideal candidate for our numerical experiments. In the following paragraphs, we summarise the potential of the TRAPPIST-1 system in general, and TRAPPIST-1e in particular, in the search for biosignatures.

The TRAPPIST-1 system has gained significant scientific attention in recent years, especially in planetary sciences and astrobiology, owing to its exceptional features. The star, with a spectral type of \(M 8.0\pm0.5\) \citep{gillonTemperateEarthsizedPlanets2016}, is known for hosting the highest number of rocky planets discovered so far. This makes it an ideal candidate for atmospheric detection, particularly as, compared to other stars within 40 pc, observations using JWST can achieve favourable signal-to-noise ratio levels (\(\text{SNR}\sim 5.5\)) within a realistic observing time (200 h) \citep{gillonTRAPPIST1JWSTCommunity2020}.

Despite being approximately 7.6 billion years old \citep{burgasserAgeTRAPPIST1System2017} and displaying moderate magnetic activity compared to other stars of its class and age \citep{gillonTRAPPIST1ItsCompact2024}, its XUV radiation emission remains significantly high, suggesting that it could potentially erode the atmospheres of the surrounding planets \citep{wheatleyStrongXUVIrradiation2017, bourrierTemporalEvolutionHighenergy2017}.

The orbital configuration of the TRAPPIST-1 system suggests that they may have migrated from farther regions, potentially leading to a rich accumulation of volatiles \citep{huangHuangShuoDynamicsTRAPPIST1System2022,agolRefiningTransittimingPhotometric2021}. This makes it an excellent laboratory for studying rocky planets orbiting ultracool dwarfs and their atmospheric retention capacity \citep{airapetianImpactSpaceWeather2020}. This is important for understanding habitability around such stars, which may host most potentially habitable planets in our galaxy \citep{gillonTRAPPIST1ItsCompact2024}.

Research on atmospheric stability suggests that these planets underwent phases of runaway greenhouse effect and intense desiccation due to stellar radiation before the star reached the main sequence \citep{lugerExtremeWaterLoss2015}. Currently, planets e, d, and f reside within the conservative habitable zone, suggesting favorable conditions for retaining secondary atmospheres. Models indicate the possible existence of atmospheres dominated by CO$_2$, such as those of Venus and Mars in the Solar System. Under more restricted conditions, their atmospheres can also be dominated by other molecular species, such as H$_2$O, O$_2$, N$_2$, CH$_4$, or NH$_3$ \citep{turbetReviewPossiblePlanetary2020}. The case of Earth is that of a planet dominated by N$_2$.

Transmission spectroscopy using the JWST instruments has been identified as the most suitable method for detecting these atmospheres, with missions already scheduled to observe transits on each planet \citep{gillonTRAPPIST1JWSTCommunity2020, turbetReviewPossiblePlanetary2020}.

Regarding the composition of these exoplanets, measurements indicate that the planets are primarily rocky, with TRAPPIST-1 e standing out for a density that makes the system closest to Earth in comparative terms, though not identical in iron proportion or volatile content. These data support the potential existence of liquid-water oceans based on the projection that TRAPPIST-1 e could have an iron fraction of 25\% \citep{agolRefiningTransittimingPhotometric2021}. TRAPPIST-1 e is of significant astrobiological interest, given its potential to present a wide range of atmospheric compositions, as the density suggests a high molecular mass atmosphere, with the possibility of maintaining liquid surfaces under various atmospheric compositions and pressures \citep{turbetReviewPossiblePlanetary2020, linDifferentiatingModernPrebiotic2021}. This potential is enhanced by the estimation that the initial water loss, due to the runaway greenhouse effect, was significantly less than that of the inner planets of the system \citep{turbetReviewPossiblePlanetary2020}.

Research into detecting Earth-like atmospheres in TRAPPIST-1 e using the JWST has motivated a variety of studies focused on specific aspects. On one hand, simulations of high-resolution spectra aim to replicate the spectral signature of Earth as seen from TRAPPIST-1 e, providing a crucial theoretical framework for identifying biosignatures \citep{kalteneggerHighresolutionTransmissionSpectra2020, linHighresolutionSpectralModels2022, lustig-yaegerEarthTransitingExoplanet2023}. On the other hand, the capabilities of JWST instruments to detect specific biosignatures face significant challenges in detecting molecules like O$_3$ and N$_2$O, which would surpass feasible observational time allocations for the telescope. Nevertheless, other studies emphasize that detecting CH$_4$, CO$_2$, and H$_2$O is viable within a limited number of transits \citep{linDifferentiatingModernPrebiotic2021, lustig-yaegerEarthTransitingExoplanet2023}. Additionally, research on the detection of stratospheric clouds concludes that, given the observational capabilities of the JWST, such detection is not yet feasible \citep{doshiStratosphericCloudsNot2022}.

In the following section, we enumerate and describe the biosignatures/bioindicators we have selected for our numerical experiments with TRAPPIST-1 e. As mentioned before, some of the biosignatures/bioindicators we select here have been identified as the most problematic to detect using the present observational capabilities and, for that reason, require novel techniques such as the one proposed here.

\section{Selected biosignatures and bioindicators}
\label{sec:biosignatures}

\subsection{\texorpdfstring{CH$_4$ and CO$_2$: primary and secondary bioindicators}{CH_{4} and CO_{2} Signals of Habitability}}\label{ch_4-and-co_2-signals-of-habitability}

Methane is common in reducing atmospheres, such as hydrogen-rich atmospheres, and may be produced under anaerobic conditions. On Earth, methane is mainly produced by bacteria and plays a crucial role in the greenhouse effect, although it also originates from non-biological sources \citep{schwietermanExoplanetBiosignaturesReview2018}. This molecule is considered a bioindicator, especially in oxidising environments because oxidation processes involving \(OH\) radicals continuously remove it, so its presence may indicate a chemical disequilibrium arising from complex processes such as life. For instance, \cite{schwietermanExoplanetBiosignaturesReview2018} have proposed that methane serves as a biosignature when O$_2$ or O$_3$ are present in an atmosphere, since these are highly oxidising molecular species. However, other molecules, such as CO$_2$, which also reflect an oxidising atmospheric state, could support the case for methane as a bioindicator.

It should be noted that CH$_4$ can also be continuously replenished on an uninhabited planet through the reaction of CO$_2$ with liquid water via serpentinisation, a process involving the hydration of ultramafic minerals such as olivine \citep{grenfellAtmosphericBiosignatures2018,schwietermanExoplanetBiosignaturesReview2018}. Therefore, while methane may not be a primary biosignature/bioindicator in this context, its presence still indicates the existence of liquid water and, consequently, habitable conditions. Thus, methane can be considered a secondary biosignature in this scenario. \cite{guzman-marmolejoAbioticProductionMethane2013} suggested that N$_2$-CO$_2$ dominated atmospheres with CH$_4$ concentrations greater than 10 ppmv could only be produced via biological processes, providing a method to distinguish between abiotic and biotic sources of methane.

\subsection{O\(_2\) and O\(_3\)}\label{sec:bio-O}

Molecular oxygen (O$_2$) on Earth is primarily the result of oxygenic photosynthesis. Historically, its detection on an exoplanet was considered a robust biosignature, as abiotic production under Earth-like conditions is minimal \citep{legerPresenceOxygenExoplanet2011}. However, recent studies have identified multiple abiotic pathways for O$_2$ and O$_3$ accumulation, prompting a reassessment of their reliability as indicators of life \citep{meadowsExoplanetBiosignaturesUnderstanding2018, schwietermanOverviewExoplanetBiosignatures2024}. These include photolysis of CO$_2$, hydrogen escape, and other mechanisms dependent on planetary and stellar conditions.

One such abiotic mechanism is CO$_2$ photolysis, which can generate O$_2$ in dry, CO$_2$-rich atmospheres. The recombination of CO and O into CO$_2$ is spin forbidden, causing a slow reaction rate and enabling O$_2$ accumulation, particularly under low humidity conditions \citep{gaoSTABILITYCO2ATMOSPHERES2015}. Additionally, on planets with surface oceans and humid atmospheres, UV radiation from M-dwarf stars can enhance CO$_2$ photolysis, contributing to O$_2$ and O$_3$ accumulation \citep{domagal-goldmanABIOTICOZONEOXYGEN2014, harmanABIOTICO2LEVELS2015}. Another key scenario involves hydrogen loss in planets that have undergone a runaway greenhouse phase. During the pre-main-sequence stage of M-dwarf stars, their intense XUV irradiation can induce massive water loss, leading to preferential hydrogen escape and leaving behind oxygen \citep{lugerExtremeWaterLoss2015}. Oceans play a critical role in this process, as extensive surface water reservoirs can provide a continuous source of H$_2$O for photodissociation. The subsequent escape of hydrogen to space results in the accumulation of O$_2$; depending on the efficiency of oxygen sinks (e.g., crustal oxidation, ion escape, reactions with volcanic gases), some planets may sustain oxygen-rich atmospheres without biological input \citep{schwietermanOverviewExoplanetBiosignatures2024}.

To differentiate biotic from abiotic O$_2$, several spectral diagnostics can be employed:

The simultaneous presence of CO and O$_2$ suggests CO$_2$ photolysis, as CO would typically be removed in a biosphere with active oxidative processes \citep{schwietermanExploringHabitabilityMarkers2016, schwietermanOverviewExoplanetBiosignatures2024}.

O$_2$–O$_2$ Collision-Induced Absorption (CIA) at 1.06, 1.27, and 6.4~$\mu$m signals high O$_2$ partial pressures, indicative of abiotic oxygen retention due to hydrogen loss \citep{fauchezSensitiveProbingExoplanetary2020}.

A low abundance of N$_2$ may suggest an atmosphere where the absence of non-condensable gases lifts the tropopause, enhancing H$_2$O photodissociation and subsequent hydrogen escape \citep{wordsworthABIOTICOXYGENDOMINATEDATMOSPHERES2014}.

On the other hand, O$_3$ is formed from O$_2$ via photochemical processes and therefore acts as a secondary biosignature or \textit{bioindicator}—its detection suggests the presence of oxygen and hence potential biological activity. However, models indicate that O$_3$ levels are highly dependent on the UV output of the host star. In cool M-dwarfs, there is a Goldilocks zone for ozone concentration that depends on the peak of UV emission \citep{grenfellSensitivityBiosignaturesEarthlike2014}. This suggests that significant amounts of O$_3$ could accumulate from lower O$_2$ levels, which may have originated from abiotic processes.

O$_3$ exhibits a strong spectral signature in the near-infrared (NIR) and mid-infrared (MIR), making it easier to detect than O$_2$ with instruments such as JWST/NIRSpec and MIRI \citep{schwietermanOverviewExoplanetBiosignatures2024}. While the most prominent absorption peak occurs in the MIR range, \cite{lustig-yaegerEarthTransitingExoplanet2023} found that the 4.7~$\mu$m peak in the NIRSpec range allows for more precise retrievals. Despite the lower O$_3$ signal at this wavelength, they highlight the importance of considering instrument sensitivity across different wavelengths, rather than solely relying on molecular absorption band strength when targeting specific molecules.

The simultaneous presence of O$_2$ (detected indirectly via O$_3$) and CH$_4$ in an exoplanet would indicate an atmosphere in redox disequilibrium (see, e.g., \citep{saganSearchLifeEarth1993}). This combination could suggest an atmosphere actively regulated by life, as these molecules react quickly to produce H$_2$O and CO$_2$ \citep{linDifferentiatingModernPrebiotic2021, schwietermanExoplanetBiosignaturesReview2018}.

\bigskip

Even though many more molecular species have been proposed as biosignatures (see, e.g., \citealt{grenfellAtmosphericBiosignatures2018,schwietermanExoplanetBiosignaturesReview2018, schwietermanOverviewExoplanetBiosignatures2024} and references therein), in our numerical experiments we will focus on CH$_4$ and O$_3$ to demonstrate the capabilities of our approach. In future works, we expect to widen the set of molecular species for studying the detectability of other potential biosignatures.

\section{Mitigation of Stellar Contamination in Transmission Spectra}\label{sec:stellar-auto}

The accurate characterisation of exoplanetary atmospheres via transmission spectroscopy is significantly impacted by stellar contamination, particularly when observing planets orbiting active stars such as M dwarfs. Addressing this challenge requires a detailed understanding of the stellar contamination phenomenon and robust computational methods to effectively remove or mitigate its effects from observed spectra. In this context, we first describe the phenomenon of stellar contamination and subsequently introduce a machine learning technique employing denoising autoencoders specifically tailored to tackle these distortions.

\subsection{Stellar contamination}\label{sec:stellar-contam}

One of the most significant obstacles when dealing with transmission planetary spectra, especially around relatively active stars, is the so-called stellar contamination. The effect, which is technically called {\it transit light source effect (TLS)}, encompasses the alterations in the transmission spectra arising from heterogeneities in the star’s photosphere \citep{rackhamTransitLightSource2018,rackhamTransitLightSource2019,rackhamEffectStellarContamination2023}. These heterogeneities are not easily revealed in photometric observations of the star and must often be modelled on a case-by-case basis \citep{rackhamRobustCorrectionsStellar2024}. 

Stellar heterogeneities, including cool starspots and hot faculae, differ from the average photosphere in both temperature and spectral output. Consequently, the light recorded during a transit includes not only the planet’s atmospheric signature but also uneven contributions from active, unocculted stellar regions. In \autoref{fig:autoencoder} we show several examples of how an original transmission spectra (first column in figure) is modified by the presence of stellar contamination (second column). As observed, stellar contamination may potentially mimic or mask genuine atmospheric features. Thus, for instance, these irregularities can hinder the detection of key molecules, such as water, especially in low-resolution observations \citep{genestEffectStellarActivity2022}. Moreover, stellar contamination can alter inferred planetary radii, inflate molecular abundance estimates, or reduce important atmospheric signals \citep{rackhamTransitLightSource2018}.

The effect of stellar contamination is particularly critical for rocky exoplanets orbiting M dwarfs, such as TRAPPIST-1 e, which is our model planet here. These stars normally exhibit high magnetic activity and hence notable photospheric heterogeneities \citep{iyerInfluenceStellarContamination2020}. In fact, stellar contamination has already been identified in transit observations conducted with JWST/NIRISS on TRAPPIST-1 b \citep{limAtmosphericReconnaissanceTRAPPIST12023}, confirming theoretical expectations.

Mitigation strategies involve employing models to correct false contamination signals or explicitly incorporating these effects into atmospheric retrieval schemes \citep{rackhamRobustCorrectionsStellar2024}. Another approach uses out-of-transit stellar spectra to constrain surface heterogeneities before correcting the resulting transmission spectrum. However, these methods face substantial challenges. Current models often fail to accurately reproduce the spectra of M dwarfs, such as TRAPPIST-1, impeding effective contamination corrections \citep{rackhamRobustCorrectionsStellar2024}. Furthermore, there is considerable degeneracy between the properties of stellar heterogeneities and planetary atmospheric characteristics, making it difficult to disentangle these contributions \citep{iyerInfluenceStellarContamination2020}. Nowadays, more accurate models and additional observations are essential to better characterise stellar inhomogeneities and their effects on transmission spectra.

For the purpose of our numerical experiments, we have devised a strategy to remove stellar contamination from our synthetic spectra using the so-called de-noising autoencoders (see below).

Contaminated transmission spectra $\left(D_{\lambda,\text{obs}}\right)$ can be mathematically expressed as \citep{rackhamTransitLightSource2018}:
\begin{equation}
 D_{\lambda,\text{obs}} = \epsilon_{\lambda}(f_\mathrm{spot},f_\mathrm{fac},T_\mathrm{spot},T_\mathrm{fac}) \, D_{\lambda}, \label{eq:stellar-contam}
\end{equation}
where \(D_{\lambda}\) is the nominal transit depth attributable solely to the planet, and the function \(\epsilon_\lambda\) encodes the stellar contamination spectrum. The function \(\epsilon_\lambda\) depends on the properties of the stellar surface heterogeneities, namely the spots and faculae. In particular, \(f_\mathrm{spot}\) and \(f_\mathrm{fac}\) denote the fraction of the stellar surface covered by spots and faculae, respectively, while \(T_\mathrm{spot}\) and \(T_\mathrm{fac}\) represent their respective temperatures. Moreover, the terms \(F_{\lambda, \text{spot}}\), \(F_{\lambda, \text{fac}}\), and \(F_{\lambda, \text{phot}}\) correspond to the spectral flux densities at wavelength \(\lambda\) for the spots, faculae, and the photosphere, respectively. According to \citealp{rackhamTransitLightSource2018} (Equation 3), a simplified expression for this contamination function is given by:
\begin{equation}
 \epsilon_{\lambda,\text{het}} = \frac{1}{1 - f_{\text{spot}} \left( 1 - \frac{F_{\lambda, \text{spot}}}{F_{\lambda, \text{phot}}} \right) - f_{\text{fac}} \left( 1 - \frac{F_{\lambda, \text{fac}}}{F_{\lambda, \text{phot}}} \right)}.
\end{equation}

\subsection{Autoencoder for denoising contaminated spectra}\label{sec:autoencoders}

Autoencoders (AE) are versatile neural networks widely used in unsupervised learning. These neural networks are used for many different tasks including, but not restricted to, dimensionality reduction, anomaly detection, and noise removal (see, e.g., \citealt{berahmandAutoencodersTheirApplications2024} and references therein).

Typically, an autoencoder consists of two main components: an encoder network that compresses the input data into a so-called reduced latent space. In spectral analysis terms, the encoder converts a spectrum with, for instance, 100 channels or features into a smaller set (e.g., 10 features) that encodes the essential information from the original spectrum. The other component, the decoder, reconstructs the data from the compressed representation. The encoder and decoder pair are trained to minimise the loss produced during the encoding-decoding process.

Among all their variants, denoising autoencoders are specifically designed to handle noisy or corrupted inputs. During training, the model receives intentionally corrupted data as input and is tasked with reconstructing the original, uncorrupted data. This process effectively isolates and removes noise while preserving essential features (see, e.g., \citealt{berahmandAutoencodersTheirApplications2024}).

In this study, we designed and developed a denoising autoencoder with a custom architecture tailored to mitigate stellar contamination and noise in transmission spectra. The model is implemented in \textit{Keras} \citep{chollet2015keras} with a \textit{TensorFlow} backend \citep{tensorflow2015-whitepaper}. Our design incorporates multiple hidden layers with Swish activation functions, dropout regularisation to reduce overfitting, and a bottleneck layer for dimensionality reduction, preserving only the most salient features. The details of our design are beyond the scope of this paper and will be presented in a forthcoming publication. Nonetheless, all source code is available alongside the programs and data accompanying this work (see \autoref{sec:data-availability}).

In \autoref{fig:autoencoder} we show several examples of applying our trained autoencoder to three representative synthetic spectra used in our numerical experiments. We can see how well our autoencoder denoises the contaminated and noisy spectra (panels in the second column in the figure), yielding clean versions (third column) that closely match the original uncontaminated spectra (first column).

The two-step approach—denoising and classification (see the next sections)—described in this paper underscores the promise of integrating state-of-the-art machine learning techniques, such as de-noising autoencoders, with established classification methods to tackle complex astrophysical problems.

\begin{figure}
    \centering
    \includesvg[width=1.0\linewidth]{images/autoencoder.svg}
    \caption{Examples of the autoencoder implementation designed to mitigate stellar contamination in three representative synthetic transmission spectra: an airless planet (first row), a planet with a CO$_2$-rich atmosphere (second row), and a planet with potential biosignatures (third row). The input spectra have an SNR of 3 and include stellar contamination levels of $f_{\mathrm{spot}} = 0.08$ and $f_{\mathrm{fac}} = 0.54$. The panels in the first column display the synthetic spectra free from noise and contamination; the second column shows the input spectra affected by noise and contamination; and the third column presents the reconstructed spectra returned by the autoencoder.}

    \label{fig:autoencoder}
\end{figure}

\section{Generation of training data}\label{sec:gen-data}

To train our algorithm, we need to generate a large sample of synthetic transmission spectra and their corresponding realizations with different levels of noise and stellar contamination. For this purpose, for generating the transmission spectra we use the {\tt TauREx 3} framework \citep{al-refaieTauRExFastDynamic2021} and, stellar contamination effects were simulated using the {\tt POSEIDON} package \citep{macdonaldHD209458bNew2017,macdonaldPOSEIDONMultidimensionalAtmospheric2023}.

Due to the complexity involved in generating a comprehensive dataset with varying mixing ratios of fill and potential biosignature gases, atmospheric temperatures, pressures, and instrumental noise, we developed an independent Python package called {\tt MultiREx}. The package is publicly available\footnote{\url{http://pypi.org/project/multirex}}. All scripts created for our numerical experiments and to generate the plots in this paper are available in the {\tt GitHub} public repository for the package\footnote{\url{https://github.com/D4san/MultiREx-public}}.

The synthetic transmission spectra were generated using the planetary properties of TRAPPIST-1 e, with the parameters provided by \citet{agolRefiningTransittimingPhotometric2021} (see \autoref{tbl:trappistvalues}). The spectrum of TRAPPIST-1 was obtained from the Phoenix model grids \citep{husserNewExtensiveLibrary2013}. The atmospheric model implemented consists of 100 layers, with an isothermal temperature profile featuring three possible temperatures: 200 K, 287 K, and 400 K, with a common base pressure of \(10^{5}\) Pa ($\approx 1$ atm) and a top pressure of \(10^{-3}\) Pa.

\begin{table}
\caption{\label{tbl:trappistvalues}Planetary and stellar parameters as provided by \citep{agolRefiningTransittimingPhotometric2021}.}

\centering
\begin{tabular}{l l}
\hline
\multicolumn{2}{c}{\textbf{TRAPPIST-1}} \\
\hline
Radius ($R_{\odot}$) & 0.0898 \\
Mass ($M_{\odot}$) & 0.1192 \\
$T_{\text{eff}}$ (K) & 2566 \\
\\
\hline
\multicolumn{2}{c}{\textbf{TRAPPIST-1 e}} \\
\hline
Radius ($R_{\oplus}$) & 0.920 \\
Mass ($M_{\oplus}$) & 0.692 \\
Semi-Major Axis (au) & 0.02925 \\
\hline
\end{tabular}
\end{table}

Although we are training our algorithms to detect planets with potential biosignatures, including atmospheres that are not necessarily conducive to life—such as those with freezing temperatures of 200 K or infernal conditions at 400 K, which are incompatible with our target model—the training of a machine-learning algorithm requires the most diverse set of conditions achievable in our models. Indeed, training the algorithms with a narrow range of temperatures, either strictly compatible with life or matching the atmospheric composition of the model planet, may lead to overfitting.

Atmospheres were created with N$_2$ as the fill gas and a fixed mixing ratio of CO$_2$ of \(10^{-2}\). This concentration was selected as an intermediate scenario between the low CO$_2$ levels observed in Earth's current atmosphere and those predicted for more primitive atmospheric conditions (see, e.g., \citealt{linDifferentiatingModernPrebiotic2021}). It should also be noted that the higher the level of CO$_2$, the harder it becomes to detect, using traditional retrieval algorithms, certain interesting biosignatures/bioindicators such as ozone. However, independent of the numerical experiments described in the following sections, we have run tests with higher concentrations of CO$_2$ and observed that, even above this level, the molecule’s effects on our results are negligible.

In addition to the fill gas, we include, depending on the case, three additional molecules: CH$_4$, O$_3$, and H$_2$O. These molecules were added such that an atmosphere could contain none, one, two, or all three simultaneously. The mixing ratio of each molecule was selected from 8 possible values, all of which are distributed log-uniformly between \(10^{-8}\) and \(10^{-1}\).

Additionally, we included in all the training sets spectra of airless planets (i.e. planets with a flat transmission spectrum). These control scenarios serve as baselines to assess the performance of classification algorithms designed to detect molecular species where no atmospheric features are present.
To account for stellar contamination, we calculated the \(\epsilon\) function (see \autoref{eq:stellar-contam}) using \texttt{POSEIDON} \citep{macdonaldHD209458bNew2017,macdonaldPOSEIDONMultidimensionalAtmospheric2023}. For this procedure, we follow the method outlined by \citet{rackhamTransitLightSource2018}, who derived spot and facula temperature parameterisations as well as coverage fractions for the TRAPPIST-1 system.

For the spot and faculae temperatures, we used fixed values of \(T_{\text{spot}} = 0.86\,T_{\text{phot}}\) and \(T_{\text{fac}} = T_{\text{phot}} + 100\,\mathrm{K}\), where \(T_{\text{phot}}\) represents the stellar photospheric temperature. For the coverage fractions, we based our samples on the values reported by \citet{rackhamTransitLightSource2018}: \(f_{\text{spot}} = 8^{+18}_{-7}\%\) and \(f_{\text{fac}} = 54^{+16}_{-46}\%\). Nine different combinations of \(f_{\text{spot}}\) and \(f_{\text{fac}}\) (minimum, mean, and maximum values for each), plus one uncontaminated scenario, were used when contaminating the synthetic transmission spectra. As a result, for each planetary spectrum (i.e. atmospheric temperature, composition, mixing ratio), we generate a total of ten possible contaminated spectra.

The spectral range for the synthesis of observational data was $\sim$0.69 $\mu$m to $\sim$5.3 $\mu$m, corresponding to the region covered by the NIRSpec PRISM (\(R=100\)), where the noise, calculated with Pandexo \citep{batalhaPandExoCommunityTool2017}, is relatively low. 

In total, 2188 synthetic spectra were generated: [8\(\times\)8\(\times\)8 (all molecules included) + 3\(\times\)8\(\times\)8 (three combinations of two molecules) + 3\(\times\)8 (atmospheres with a single species) + 1 (atmosphere with only fill gases)] \(\times\) 3 (temperature values) + 1 (flat spectrum). By applying the ten stellar-contamination scenarios (nine contaminated + one uncontaminated) to each of these base spectra, the overall dataset was expanded to \(21,880\) distinct synthetic spectra.

In the spectra synthesis, contributions due to Rayleigh scattering and molecular absorption were considered. For the opacities of CO$_2$, CH$_4$, and H$_2$O, we used data from the ExoMol database \citep{chubbExoMolOPDatabaseCross2021}, while the opacities of O$_3$ and N$_2$ were taken from Exo-Transmit \citep{lupuATMOSPHERESEARTHLIKEPLANETS2014, freedmanGASEOUSMEANOPACITIES2014, freedmanLineMeanOpacities2008}. 

All these molecules are present in Earth's current atmosphere and significantly contribute to the terrestrial spectrum. The combination of O$_3$ and CH$_4$ is considered a strong biosignature (see \autoref{sec:biosignatures}). Additionally, both CH$_4$ and H$_2$O are also present in prebiotic scenarios \citep{linDifferentiatingModernPrebiotic2021}.

O$_2$ was not included in the model because it does not exhibit a strong signal in the range of interest, and the CIA O$_2$-O$_2$ (which would be the major spectral signal of O$_2$) would not be detectable in the NIRSpec range with a reasonable number of transits \citep{fauchezSensitiveProbingExoplanetary2020}. 

The presence of clouds was also not included. Since they are located in the lower part of the atmosphere, they should not significantly affect the model's predictions \citep{linDifferentiatingModernPrebiotic2021, barstowHabitableWorldsJWST2016}. Additionally, according to \citet{doshiStratosphericCloudsNot2022}, the presence of stratospheric clouds would not affect transmission spectra, at least with the JWST instruments.

To train the algorithms, we added to the theoretical spectra a constant Gaussian noise across the entire wavelength range. The noise was calculated according to the assumed SNR value. As a result, the number of synthetic spectra is substantially increased with respect to the 21,880 theoretical set. In some cases by a factor of 100 or more. 

We trained the algorithms using five discrete SNR values: noise-less (denoted 0), 1, 3, 6, and 10. Larger SNR values were not used since from our {\tt PandExo} analysis, such a noisy spectra would require more than 100 transits with NIRSpec PRISM (see \autoref{sec:SNRtransits}. 

For each set of atmospheric parameters, we determined the noise standard deviation from the signal of the uncontaminated spectrum, thus ensuring consistency among the contaminated versions of that family. Finally, all spectra were normalised between their minimum and maximum values for training. For testing, however, we used a fine-grained interval of SNR.

\begin{figure*}
    \centering
    \includegraphics[width=1\linewidth]{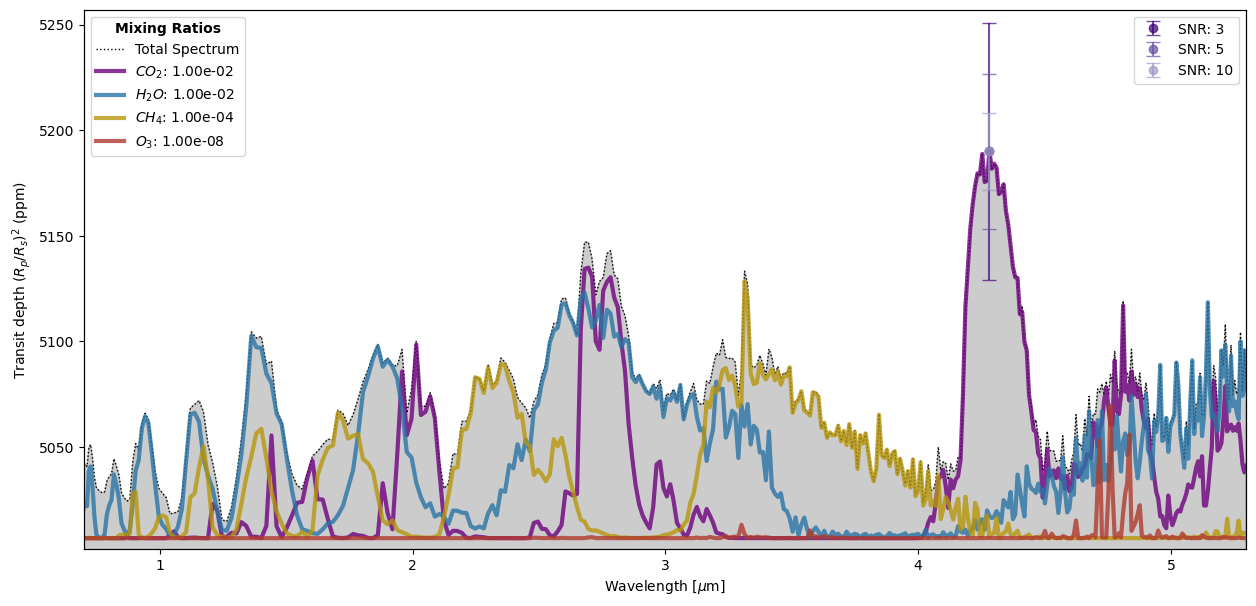}
    \caption{Absorption contributions from different molecules in a potentially inhabited TRAPPIST-1 e spectrum with surface concentrations similar to a Neoproterozoic Earth \citep{kalteneggerHighresolutionTransmissionSpectra2020} and a mixing ratio of H$_2$O fixed at \(10^{-2}\). Each colour represents a molecule (see legend). Error bars illustrate noise based on the SNR, with the strongest CO$_2$ peak used to define the signal. The error bars represent a $1\sigma$ uncertainty.}

    \label{fig:spec-dataset}
\end{figure*}

In \autoref{fig:spec-dataset}, we show a representative transmission spectrum containing all interesting chemical species in concentrations similar to those of a Neoproterozoic Earth \citep{kalteneggerHighresolutionTransmissionSpectra2020}. We can see that many peaks may become obscured by other molecules if the mixing ratio and noise level of the confounding molecule are increased.

To illustrate the strength of the noise in the simulated transmission spectrum in \autoref{fig:spec-dataset}, we include a bar over the strongest CO$_2$ peak, indicating the magnitude of the relative error. As expected for an SNR of 3, the noise can be as high as one third of the peak amplitude.

For illustrating the effect of the level of noise in the spectra we deal with in this work, we show in \autoref{fig:noisy-spectra} several examples of signals at SNR=1. These figures illustrate well how featureless noisy spectra seem even at large mixing ratios of the interesting molecules.

\begin{figure}
\centering    \includesvg[width=1\linewidth]{images/noisy_spectra.svg}
\caption{Examples of synthetic transmission spectra for a TRAPPIST-1 e planet under noisy conditions at \(\text{SNR} = 1\). Each panel corresponds to a distinct combination of methane (CH$_4$) and ozone (O$_3$) mixing ratios, illustrating how noise at this level affects the observable spectral features. The top three panels (yellow, red, and blue points) show the noisy flux measurements across the wavelength range for three different CH$_4$--O$_3$ scenarios. The bottom panel compares these scenarios in a single line plot, highlighting the variation in spectral peaks and the extent to which noise can obscure molecular signatures.}

\label{fig:noisy-spectra}
\end{figure}

\section{Biosignature Random Forest Classification}\label{sec:RF}

Various classification experiments employing Random Forest (RF) were performed to investigate the potential for biosignature detection:

\begin{enumerate}
    \item The first and simpler experiment aimed to identify ``interesting'' targets, i.e.\ planets that could potentially have biosignatures/bioindicators and thus merit further investigation. We refer to this experiment and the resulting model as the \textbf{\em Binary Classifier} (BC).

    \item The second experiment is more complex. It seeks to classify planets by specific molecules (e.g.\ determining whether a planet could contain ozone). We call this the \textbf{\em Multilabel Classifier} (MC).

    \item The third experiment involves training what we call ``specialist'' algorithms. These algorithms are trained on planets that have a specific biosignature molecule, yet they are applied to classify the entire synthetic spectra dataset. We refer to this as the \textbf{\em Specialised Classifier} (SC).
\end{enumerate}

For the implementation and performance evaluation of the RFs, we used the diverse functionalities provided by {\tt scikit-learn} \citep{pedregosaScikitlearnMachineLearning2011}. This approach enabled a comprehensive analysis of biosignatures/bioindicators, leveraging the robust machine learning tools available within that library.

A critical step in RF classification is determining the optimal threshold for classifying a planet as positive or negative. Although a conventional threshold of 0.5 is typically used, we adjusted it to increase the Recall, given the significant need to distinguish between FP and FN. This adjustment initially led to an increase in FP, prompting further lowering of the threshold. The threshold was optimised to maintain the TNR above 0.6 across all SNR levels.

\subsection{Binary Classifier}\label{sec:binaryclass}

As a first experiment, we generated a large set of noisy synthetic spectra encompassing all possible compositional combinations: only fill gases, only CH$_4$, only O$_3$, combinations of CH$_4$ and O$_3$, only H$_2$O, combinations of H$_2$O and CH$_4$, and so on. Each planet was labelled as {\tt bio} (True or 1) if it had at least one biosignature, and {\tt non-bio} (False or 0) otherwise. 

This general experiment required the broadest range of atmospheric compositions among all performed experiments The distribution of planets having different combinations of biosignatures/bioindicators is summarised in 
\autoref{tbl:data-interesting}. As we explained before, we generate synthetic spectra with 6 levels of SNR (rows in the table). The binary codes at the head of the table indicate the presence (1) or absence (0) of CH$_4$, O$_3$, and H$_2$O, respectively. For instance, the third column, labelled 010, corresponds to planets without CH$_4$ (0), containing O$_3$ (1), and in the absence of H$_2$O (0). The values in each cell represent the total number of noisy spectra generated for each atmospheric type.

\begin{table}
  \begin{center}
    \caption{\label{tbl:data-interesting}Distribution of training data for the RF classifier targeting interesting planets. Each row represents a different SNR level (with 0 indicating a noise-free spectrum), while each column is labelled with a three-digit binary code indicating the presence (1) or absence (0) of CH$_4$, O$_3$, and H$_2$O, respectively (e.g., “010” means no CH$_4$, presence of O$_3$, and no H$_2$O). The values in each cell (in thousands) denote the total number of noisy spectra generated for that specific atmospheric configuration.}

    \resizebox{\linewidth}{!}{%
      \begin{tabular}{ccccccccccc}
        \hline 
        \textbf{SNR} & \textbf{000} & \textbf{100} & \textbf{010} & \textbf{001} & \textbf{110} & \textbf{101} & \textbf{011} & \textbf{111} & \textbf{airless} \\
        \hline 
        1  & 300 & 60   & 60   & 60   & 57.6  & 57.6  & 57.6  & 76.8  & 100 \\
        3  & 300 & 60   & 60   & 60   & 57.6  & 57.6  & 57.6  & 76.8  & 100 \\
        6  & 300 & 24   & 24   & 24   & 38.4  & 38.4  & 38.4  & 46.08 & 100 \\
        10 & 300 & 24   & 24   & 24   & 38.4  & 38.4  & 38.4  & 46.08 & 100 \\
        0  & 300 & 60   & 60   & 60   & 96    & 96    & 96    & 57.6  & 100 \\
        \hline
      \end{tabular}%
    }
  \end{center}
\end{table}

In the example above, to train the algorithm using synthetic spectra with 010 compositions and an SNR of 10 (fourth row), we generated 250 noisy spectra for each O$_3$ concentration value (8 in total) and for each atmospheric temperature (3 in total). Considering the 10 different contamination scenarios (9 contaminated + 1 uncontaminated), this resulted in 2500 spectra per O$_3$ concentration and temperature combination, leading to a total of 60,000 spectra for this composition alone. Similarly, for the case of planets having all molecules (111), we generated 5 noisy spectra per molecule concentration (8$\times$8$\times$8), per temperature value (3), and per contamination scenario (10 in total), resulting in 76,800 spectra for this composition.

A total of 3,770,560 synthetic spectra were used in this experiment.

To avoid overfitting on the highest SNR spectra, we provided the algorithm with a smaller number of samples having SNR values of 6 and 10.

It should be emphasised that the quantity and distribution of data labelled as {\tt bio} or {\tt non-bio} can significantly influence RF performance, especially for low SNR signals. Adjusting the training data proportion towards more negative or positive examples, as necessary, can enhance the model's ability to recognise the presence or absence of certain molecules under challenging noisy conditions.

Using this data, we trained an RF classifier to determine whether new spectra are {\tt interesting} or {\tt not-interesting}. In this context, an interesting spectrum is defined by the potential presence of CH$_4$ or O$_3$, while the presence of H$_2$O is treated as interference rather than as a biosignature. The training was conducted using 80\% of the data, with an initial validation performed on the remaining 20\%.

Our RF model contained 100 estimators, used the entropy criterion, and a minimum samples leaf of 3. After training, the classification threshold was set at 0.45.

The resulting trained RF was tested using a set of approximately 900,000 new synthetic spectra per SNR value, equally distributed between the {\tt bio} and {\tt non-bio} categories. The {\tt non-bio} spectra consisted of 150,000 planets with CO$_2$ and N$_2$, 150,000 planets without an atmosphere, and approximately 150,000 with H$_2$O, CO$_2$, and N$_2$. The {\tt bio} spectra comprised roughly 75,000 samples for each possible combination of biosignature/bioindicator molecules (6 combinations), giving a total of approximately 450,000 test spectra. As can be seen, the test set is balanced between samples with and without biosignature/bioindicator molecules.

\subsubsection{BC metrics}
\label{uusec:BC-metrics}

In \autoref{fig:interesting-gmetrix}, we show the value of each metric defined in \autoref{sec:metrics} as a function of SNR for the trained algorithm. It is important to stress that these results are obtained when classifying spectra with all possible mixing ratios. Therefore, this metric provides a conservative measure of the algorithm's performance. If we were to focus only on the planets with the highest concentrations of the biosignature/bioindicator gases, the results would be much better.

As expected, as SNR increases, most metrics improve. Even in the worst-case scenario of SNR = 1, all metrics have a value above 0.6, indicating that even with the noisiest and most contaminated spectra (as represented by the shaded area), the algorithm performs better than a blind guess.

\begin{figure}
    \centering
    \includesvg[width=1.07\linewidth]{images/auto-BC-globalmetrics.svg}
    \caption{Global metrics (TNR, Recall, Precision, and F1 score; see \autoref{tab:metrics} for definitions) for the binary classification of {\tt interesting} (with biosignatures/bioindicators) and {\tt non-interesting} planets across different SNR values. The pink shaded area below 0.6 in the graphs indicates suboptimal classification performance. The solid lines represent the results for spectra without stellar contamination, while the coloured shaded areas around each curve indicate the variance in the results produced by stellar contamination.}

    \label{fig:interesting-gmetrix}
\end{figure}

Notably, the algorithm can classify {\tt non-bio} as {\tt non-interesting} planets very efficiently (TNR $\approx 0.9$) even for values of SNR as low as 3. This means that the algorithm could be very good at avoiding wasting time on uninteresting planets. This is confirmed by the behaviour of the Precision metric.

The analysis of these metrics indicates that our BC algorithm performs adequately in labelling planets with biosignatures/bioindicators as {\tt interesting}, but it performs even better at discarding planets without these features as {\tt non-interesting}. This capability is important for preventing the unnecessary allocation of research time to targets that are unlikely to yield valuable biosignature information.

\subsubsection{BC minimum mixing ratio for detection}
\label{uusec:BC-minmix}

As a secondary analysis, for each SNR value we sorted the classified synthetic spectra according to the CH$_4$ mixing ratio. We then selected the spectra corresponding to a specific CH$_4$ mixing ratio and computed our binary classification metrics solely for planets with that concentration. It is important to note that while \autoref{fig:interesting-gmetrix} presents the overall performance, it does not reveal how the classifier behaves at lower mixing ratios, as scenarios with high mixing ratios can mask the performance at the lower end.

For low mixing ratios, the Recall metric was generally low, consistent with expectations. We repeated this procedure until the Recall reached a value of 0.6; the corresponding mixing ratio is termed the \emph{minimum mixing ratio for detection}. We performed this analysis for O$_3$ as well, thereby providing a clearer insight into the classifier’s sensitivity at lower mixing ratio values.

In \autoref{fig:interesting-minmix}, we show the minimum mixing ratio required for the detection of each biosignature as a function of SNR. In the case of CH$_4$, it is observed that the algorithm, at SNRs of 1–2, successfully classifies as interesting atmospheres with mixing ratios as low as \(10^{-6}\) to \(10^{-4}\). For comparison, the present Earth has a CH$_4$ mixing ratio on the order of \(10^{-6}\), but in the past it has reached levels as high as \(10^{-3}\) or greater \citep{kalteneggerSpectralEvolutionEarthlike2007, kalteneggerHighresolutionTransmissionSpectra2020}. Notably, stellar contamination (indicated by the shaded areas in the figure) tends to lower the minimum mixing ratio required for detection. In other words, when a planetary transmission spectrum is heavily contaminated, our BC algorithm is more inclined to flag it as interesting. This effect may arise because contamination—especially from spots—can enhance certain spectral features, thereby aiding the algorithm in identifying the relevant molecules. Consequently, the presence of spots might also influence the autoencoder's ability to reconstruct an uncontaminated spectrum, ultimately improving the BC algorithm's results.

\begin{figure}
    \centering
    \includesvg[width=1.07\linewidth]{images/auto-BC-minmix.svg}
    \caption{Minimum mixing ratio required for detection as interesting (with Recall $>$ 60\%) across different SNR values.}
    \label{fig:interesting-minmix}
\end{figure}

The case of O$_3$ is particularly interesting. Our BC algorithm can identify as interesting most planets with ozone levels on the order of those observed on present-day Earth (\(10^{-5}\)–\(10^{-6}\)), starting at an SNR of 2. In the case of TRAPPIST-1 e, achieving such low SNR values would require only about 6 transits with JWST. It is important to note that although present-day ozone levels at Earth’s surface are around \(10^{-8}\), transmission spectra sample an air column that includes layers with much higher ozone concentrations, effectively raising the observed value to \(10^{-5}\)–\(10^{-6}\) (see Table 2 in \citealt{lustig-yaegerEarthTransitingExoplanet2023} and Figure 5 in \citealt{linDifferentiatingModernPrebiotic2021}).

Finally, we observe that even when the levels of O$_3$ or CH$_4$ are relatively low in inhabited planets—i.e. planets exhibiting a non-equilibrium abundance of both molecular species—our algorithm still flags such planets as interesting, even if the detectability of one of the molecules falls below the minimum mixing ratio threshold. Conversely, if a planet exhibits only one of these molecules (and thus should be considered as having a bioindicator rather than a full biosignature), the BC algorithm will still label it as interesting. This observation motivates the development of multilabel algorithms capable of labelling the atmosphere according to individual species (see \autoref{uusec:MC-mulmetric} and \autoref{sec:SC-multilabel}).

\subsubsection{BC interference analysis}
\label{uusec:BC-interference}

We can use similar methods as before to evaluate the performance of the algorithm in classifying {\tt non-bio} planets with H$_2$O as interesting cases. This analysis measures the tendency to misclassifying planets due to the presence of water. As shown in \autoref{fig:spec-dataset}, the water absorption bands in the spectral range under consideration overlap significantly with most of the ozone bands and some of the methane bands. Consequently, when a planet exhibits a high water mixing ratio and a low SNR, it may be incorrectly classified as interesting.

\begin{figure}
    \centering
    \includegraphics[width=1\linewidth]{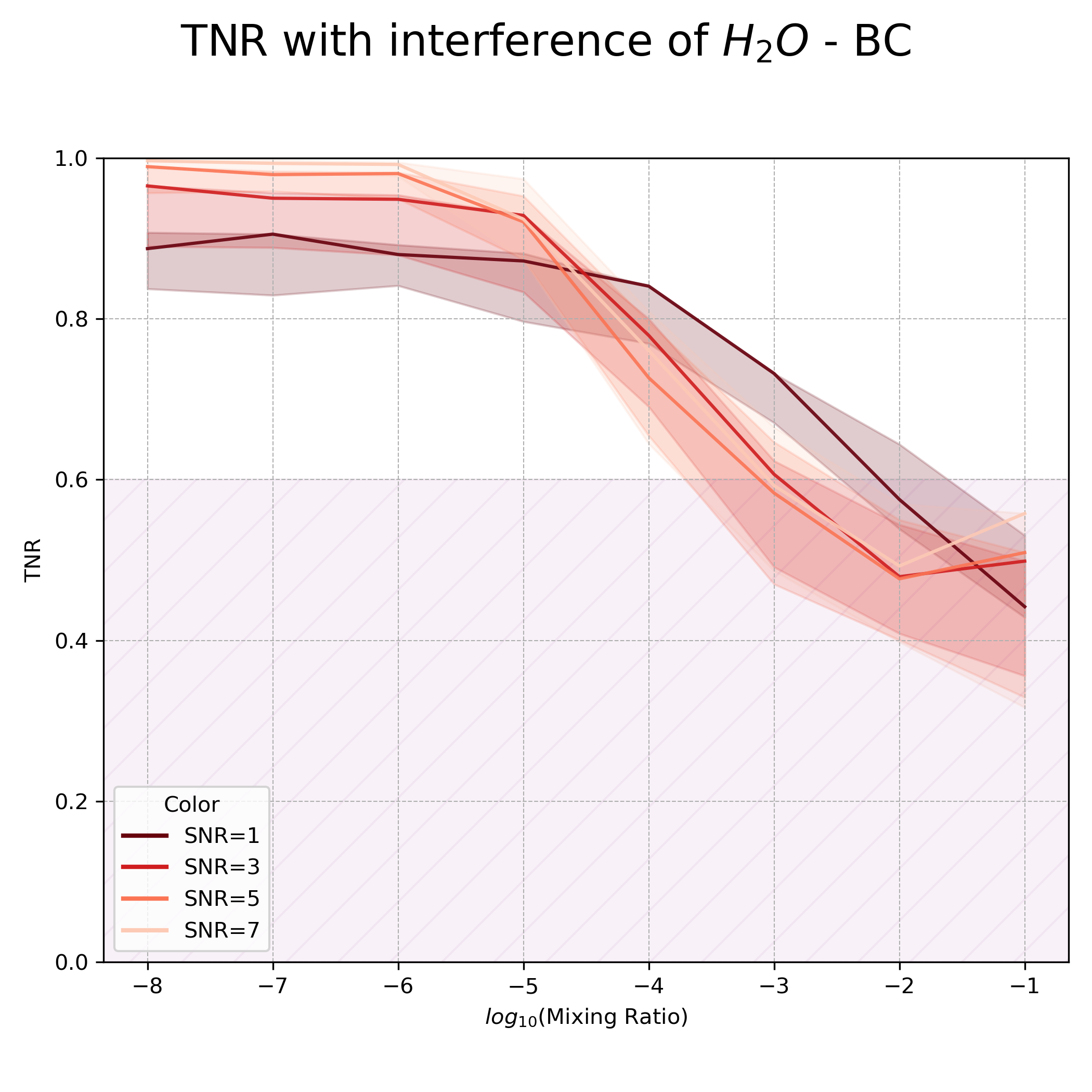}
    \caption{TNR in BC across H$_2$O mixing ratios evaluated at representative SNR values. The shaded area below 0.6 indicates suboptimal classification performance.}
    \label{fig:TNR-water}
\end{figure}

In \autoref{fig:TNR-water}, we present the TNR (the most appropriate metric for this analysis) as a function of the water mixing ratio. Recall that a high TNR indicates that the algorithm correctly classifies the spectrum of an uninhabited planet as non-interesting. As the figure demonstrates, the TNR decreases as the water mixing ratio increases, which is consistent with the expectation that substantial water interference can lead to misclassification. Conversely, as the SNR increases, the algorithm's performance improves (i.e. TNR increases), reflecting reduced sensitivity to water interference.

We also analyze the performance of the algorithm with planets having two molecules. For that purpose, we picked samples of spectra having pairs of mixing ratios such as (CH$_4$, O$_3$): \((10^{-8},10^{-8})\), \((10^{-8},10^{-7})\), \(\ldots\), \((10^{-7},10^{-8})\), \((10^{-7},10^{-7})\), \(\ldots\). For each sample, we computed the Recall. In the columns of \autoref{fig:RecallMaps} we show the result as a heat map of this metric as a function of mixing ratio combinations. We call these \emph{Recall maps}. 

\begin{figure*}
    \centering
    \includesvg[width=1.07\linewidth]{images/auto-BC_RecallMap_SNR.svg}
    \caption{Grid of Recall maps for BC at two representative SNR levels (2 and 10). Recall was calculated for different pairs of molecules across varying mixing ratios.}
    \label{fig:RecallMaps}
\end{figure*}

It is important to point out that for these plots we used in the test set spectra with all levels of stellar contamination. That is, while in \autoref{fig:TNR-water} we classified spectra according to various levels of stellar contamination, for the nature of this 2D heat map we decided to combine all spectra irrespective of the contamination level.

We observe that even at low mixing ratios of one of the biosignatures/bioindicators, Recall can be high if the mixing ratio of the other is high enough. For example, at SNR = 2 (first row in \autoref{fig:RecallMaps}) and at a very low concentration of ozone (e.g. \(10^{-8}\) in the first panel), the presence of abundant methane (e.g. \(10^{-3}\)) causes the algorithm to classify as interesting most planets that exhibit both biosignatures/bioindicators. Our interpretation is that the algorithm learns the spectral bands of each biosignature/bioindicator independently and is able to recognise that a planet is interesting by identifying any one of them, even if the other is absent.

Additionally, as SNR increases, the algorithm becomes more conservative (yielding lower Recall values) when both mixing ratios are low, but achieves better performance (higher Recall values) at higher mixing ratios.

We repeated the same procedure using combinations of one biosignature/bioindicator and water. The resulting Recall maps are shown in the second and third columns of \autoref{fig:RecallMaps}. Although water tends to interfere with the detection of biosignatures/bioindicators, the fact that the frontier between low and high Recall areas in the Recall maps is nearly horizontal indicates that the performance of the algorithm is mainly determined by the mixing ratio of the biosignature. In other words, H$_2$O only negatively affects the Recall when the mixing ratio of the biosignature is very low.

\subsubsection{BC for airless and CO$_2$ rich planets}
\label{uusec:BC-airless}

To evaluate the robustness of the binary classifier against false positives, its performance was investigated in two critical scenarios: (i) airless planets, where noise and contamination may lead to the misinterpretation of spectral features as molecular bands when no atmosphere is actually present; and (ii) atmospheres composed of fill gases, such as CO$_2$ and N$_2$, for which spectral bands fall in the same region as the sought bioindicator—particularly in the case of O$_3$—and this “virtual overlapping” could result in erroneous classifications.

In the airless scenario, the TNR was always larger than 0.9 even at an SNR of 1 and under the most adverse stellar contamination. This means that if we take 100 test spectra of airless planets, more than 90 will be correctly classified as non-interesting. By contrast, for pure fill gas atmospheres, the TNR is approximately 0.8 at an SNR of 1, improving to values exceeding 0.9 at an SNR of 3.

These findings demonstrate that the binary classifier is robust at avoiding false positives.

\subsection{Multilabel Classifier}
\label{sec:MC}

In our previous experiment we binary classified the planets as {\tt interesting} and {\tt not-interesting}. We now aim to improve the method by introducing multiple labels intended to identify individual biosignatures/bioindicators and other interesting molecules.

For that purpose, we specifically define three labels: {\tt methane}, {\tt ozone} and {\tt water}, each with two possible values: 0 (the molecule is absent) and 1 (the molecule is present).
Focusing on the detection of each molecule individually is valuable because, as discussed in \autoref{sec:biosignatures}, different biosignatures/bioindicators have varying degrees of robustness and implications. By independently classifying each biosignature, we can better understand the presence and relevance of each molecule, thereby improving our ability to assess the potential habitability of exoplanets.

The MC random forest uses the same hyperparameters as the BC algorithm described in \autoref{sec:binaryclass} (100 estimators and a maximum depth of 200).

The training spectra used were of the same types as in the previous experiments, but the distribution varied. In \autoref{tbl:data-MC}, we show the number of training spectra according to composition and SNR for the MC dataset. From this dataset, 80\% was used for training, and an initial validation was performed with the remaining 20\%.

\begin{table}
\begin{center}
\caption{\label{tbl:data-MC}Training data distribution for the multilabel classifier (MC). This table has the same structure as \autoref{tbl:data-interesting}, with each cell's value (\textbf{in thousands of samples}) representing the total number of noisy spectra generated for that atmospheric configuration.}
    \resizebox{\linewidth}{!}{%
    \begin{tabular}{cccccccccc}
    \hline \text { \textbf{SNR} } & \textbf{000} & \textbf{100} & \textbf{010} & \textbf{001} & \textbf{110} & \textbf{101} & \textbf{011} & \textbf{111} & airless\\
    \hline 1 & 300 & 60 & 60 & 60 & 57.6 & 57.6 & 57.6 & 76.8 & 100 \\
    3 & 300 & 60 & 60 & 60 & 57.6 & 57.6 & 57.6 & 76.8 & 100  \\
    6 & 300 & 24 & 24 & 24 & 38.4 & 38.4 & 38.4 & 46.08 & 100\\
    10 & 300 & 24 & 24 & 24 & 38.4 & 38.4 & 38.4 & 46.08 & 100 \\
    0 & 300 & 60 & 60 & 60 & 96 & 96 & 96 & 76.8 & 100 \\
    \hline
    \end{tabular}
    }
\end{center}
\end{table}

After training, the classification thresholds for CH$_4$, O$_3$, and H$_2$O were set at 0.28, 0.24, and 0.36, respectively.

The model was tested by generating approximately 750,000 noisy spectra per SNR. This included 150,000 spectra for the scenario without interesting molecules (comprising only filler gases and airless cases, in equal proportions), and another 150,000 spectra for the cases where all three molecules are present. For the remaining combinations, we generated 75,000 spectra per combination, resulting in a total of 450,000 spectra. This distribution ensures a balanced test dataset, meaning that for each molecule there is an equal number of spectra where it is present and absent.

\subsubsection{MC multilabel metrics}
\label{uusec:MC-mulmetric}

In \autoref{fig:MC-mulmetric}, we show the algorithm's overall performance as measured with all the multilabel metrics de define in \autoref{subsec:multilabel}. 

\begin{figure}
    \centering
    \includesvg[width=1\linewidth]{images/auto-MC-mulmetrics.svg}
    \caption{Multilabel classification metrics for the MC as a function of SNR. For the meaning of the shaded regions, see \autoref{fig:interesting-gmetrix}. \label{fig:MC-mulmetric}}
\end{figure}

As we see, the overall performance of the algorithm is very good for most metrics even at an SNR of 1, and it improves further as the SNR increases. The only metric that performs poorly is the Perfect Match, which measures the algorithm's capacity to correctly classify the composition of a planet's atmosphere with the perfect label—irrespective of the combination of molecular species (see \autoref{tab:metrics}). For the MC algorithm, the Perfect Match does not exceed 50\% and is only achieved when the SNR is 10 or higher.

Although this result may seem disappointing, it is important to note that this metric is extremely stringent; thus, scores on the order of 0.3 are not as adverse as they might appear.

In this figure, we have also included a special Recall metric, termed \emph{Chemical Disequilibrium Recall}, which measures the algorithm's capacity to label a planet as having both CH$_4$ and O$_3$ simultaneously. This is an interesting case since the presence of both gases indicates a chemical disequilibrium—a condition recognised as one of the strongest evidences for an inhabited planet (see \autoref{sec:biosignatures}). Notably, the MC algorithm is able to flag more than 70\% of the inhabited planets in our test set as interesting cases for follow-up observations, even at an SNR of 1.

To illustrate this point about how impressive this achievement is, please refer to \autoref{fig:noisy-spectra}, which shows examples of the types of spectra we are dealing with. It is particularly noteworthy that the MC algorithm achieves accurate classification of most spectra, even under conditions of very low spectral feature contrast, which few retrieval algorithms would reliably fit.


\subsubsection{MC molecule metrics}
\label{uusec:MC-molmetric}

The novel feature of MC with respect to BC is its ability to allow for per-molecule classification. In \autoref{fig:MC-molmetric}, we show the binary metrics for the algorithm when the planets are classified by the presence of each bioindicator separately. To calculate the metric, we first select, from the training set, the spectra with a specific SNR value and containing a given molecule (e.g. O$_3$). Then, using the corresponding label (e.g. {\tt ozone}, which can have a value of 1 if the planet has ozone or 0 otherwise) and the class assigned by the algorithm, we compute the metrics. Naturally, the higher the score achieved for a given molecule, the better the algorithm is at identifying that molecule among others that may be present in the spectrum.

\begin{figure}
    \centering
    \includesvg[width=1\linewidth]{images/auto-MC-molmetrics.svg}
    \caption{Molecule-specific metrics for the MC. The shaded regions indicate the variance in performance (see \autoref{fig:interesting-gmetrix} for details).}
    \label{fig:MC-molmetric}
\end{figure}

In all metrics, better performance is attained when classifying O$_3$ (red continuous line and shaded strip) compared to CH$_4$ and H$_2$O (blue strip). Both CH$_4$ and O$_3$ exhibit a Recall above 60\% even at low SNR.

Comparing all metrics, we observe that the MC algorithm is effective at identifying interesting candidates. This is evident from the Recall (first row, second column in \autoref{fig:MC-molmetric}), which reaches values above 80\% for SNRs greater than 3. Furthermore, the TNR and Precision metrics—which help avoid wasting time on uninteresting targets—are also satisfactory. In fact, for the interesting bioindicators, even at low SNR values of 3–4, these metrics are close to or exceed 0.8, depending on the level of stellar contamination (as indicated by the shaded areas around the continuous lines).

\subsubsection{MC minimum mixing ratio for detection}
\label{uusec:MC-minmix}

Using the Recall metric, we repeat the experiment in \autoref{uusec:BC-minmix} to determine the minimum mixing ratio at which the MC algorithm can tag the presence of each molecule. The results are presented in \autoref{fig:MC-minmix}.

\begin{figure}
    \centering
    \includesvg[width=1.07\linewidth]{images/auto-MC-minmix.svg}
    \caption{Minimum mixing ratio required for detection of each molecule (with Recall > 60\%) by MC as a function of SNR.}
    \label{fig:MC-minmix}
\end{figure}

As expected, the minimum detectable mixing ratio decreases with increasing SNR, even for H$_2$O. However, the minimum mixing ratio required for the detection of water is higher than that for the biosignatures/bioindicators. This is because, in the studied spectral range, the absorption strength of water is much lower than that of the other molecules.

\subsubsection{MC interference analysis}
\label{uusec:MC-interference}

Now, we aim to evaluate the effect of one molecule's presence on causing the misclassification of a planet as if it contains another molecule. For instance, at certain mixing ratios, the presence of ozone causes the algorithm to misclassifying a planet as if it contains methane, even though methane is absent. This misclassification interference is measured by the TNR metric. We already tested this phenomenon in \autoref{uusec:BC-interference} for the BC algorithm using water as the interfering molecule.

\begin{figure*}
    \centering
    \includegraphics[width=1\linewidth]{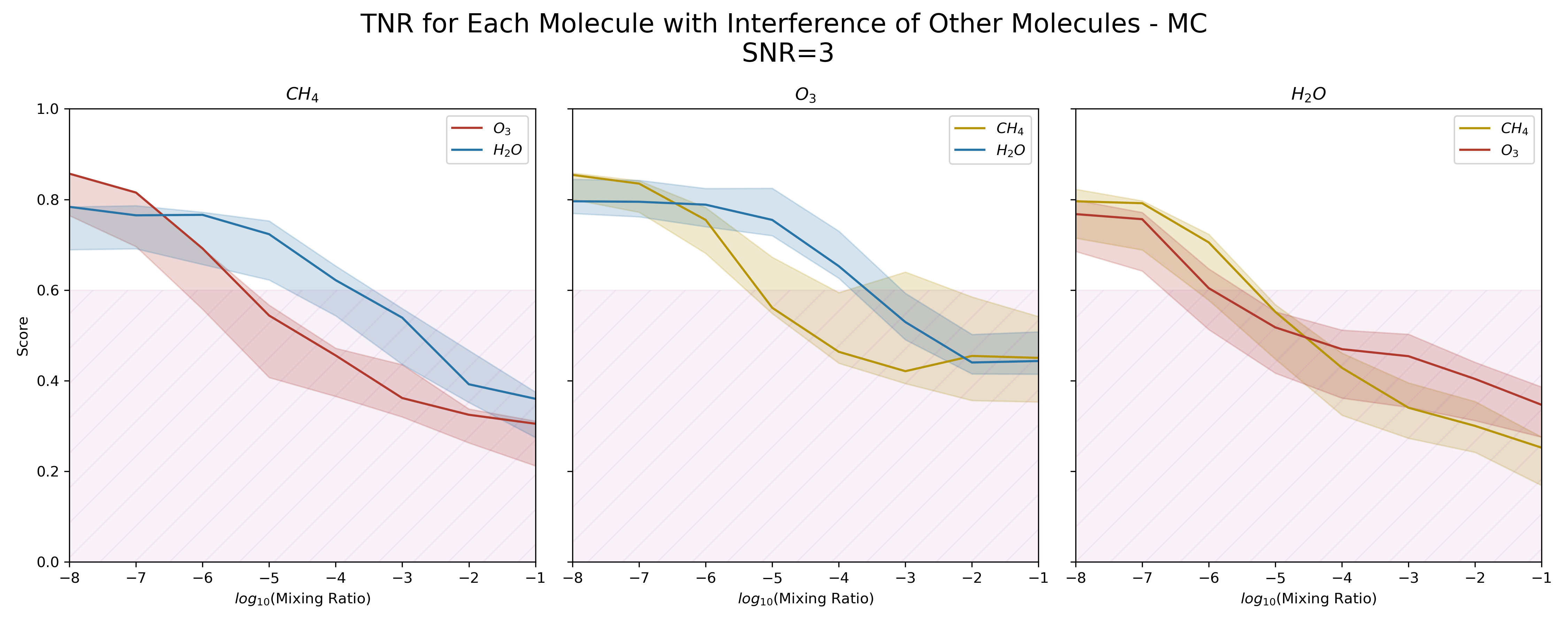}
    \caption{TNR in MC as a function of the mixing ratios for each interfering molecule, at a constant SNR. Each column represents the true negative rate (TNR) for the classification of: CH$_4$ (left), O$_3$ (center), and H$_2$O (right). For the meaning of the shaded regions, see \autoref{fig:interesting-gmetrix}. 
    }
    \label{fig:MC-interference}
\end{figure*}

In \autoref{fig:MC-interference}, we present the results of comparing the TNR measured individually for different molecules (in separate panels) while varying the mixing ratio of the interfering molecules (represented by different curves in each panel). We focus on the case for SNR = 3, as our numerical experiments indicate that the interference effect is most pronounced at this SNR value.

In all cases, increasing the mixing ratio of the interfering molecule reduced the TNR for each biosignature/bioindicator. In other words, the algorithm’s ability to correctly label as non-interesting those planets that lack a given molecule is poorest when an interfering molecule is present. The most significant effect is observed when ozone interferes with the classification of methane (first panel in \autoref{fig:MC-interference}). At an ozone mixing ratio as low as \(10^{-6}\)–\(10^{-5}\), the algorithm tends to systematically classify planets that do not contain methane as if they did. However, the TNR for O$_3$ (second panel in \autoref{fig:MC-interference}) does not decline as much at higher CH$_4$ concentrations. In contrast, the classification of H$_2$O is most affected by the presence of both biosignatures/bioindicators, leading to confusion across a range of concentrations.

It is interesting to note that the presence of water creates the least interference among all molecules. Only at very high water concentrations (exceeding approximately \(10^{-4}\)) does interference with biosignatures/bioindicators become significant. This implies that the algorithm may erroneously tag wet planets as inhabited—i.e. as possessing biosignatures/bioindicators—even if they do not.

In summary, the MC algorithm faces challenges in classifying molecules individually due to interference from other molecules. These effects are particularly concerning when the mixing ratios of the interfering molecules are relatively high. Interference between ozone and methane is less concerning, since both are considered biosignatures/bioindicators; however, interference involving water—which is not regarded as a biosignature—can be misleading. Nonetheless, the water mixing ratios required for such misclassifications are relatively large\footnote{For present-day Earth, the integrated water mixing ratio is around \(10^{-5}\) \citep{lustig-yaegerEarthTransitingExoplanet2023}, depending on the atmospheric model}.

A final, but equally important, experiment we performed with the MC algorithm was to analyse how the presence of other molecules affects the Recall in the classification of a molecule of interest. In \autoref{fig:MC-RecallMaps}, we present the Recall maps for the molecules analysed by our algorithm. This figure is analogous to \autoref{fig:RecallMaps}; however, in contrast to the latter, a single SNR value of 3 was used to illustrate the interference effect.

\begin{figure*}
    \centering
    \includesvg[width=0.9\linewidth]{images/auto-MC-RecallMap.svg}
    \caption{Grid of Recall maps for MC at SNR=3. Recall was calculated for different pairs of molecules across varying mixing ratios. Each row represents the classification for a specific molecule (CH\(_4\), O\(_3\), and H\(_2\)O, from top to bottom), with interference from other molecules as indicated in the labels.}
    \label{fig:MC-RecallMaps}
\end{figure*}

In all cases, contrary to what might be expected, interference may actually enhance the detection of a given molecule. This is reflected in the figure as darker areas in the lower-left corner of each panel. This effect is particularly pronounced for methane; even at very low methane mixing ratios, the presence of high concentrations of ozone can cause the algorithm to label a planet as having methane (as seen in the lower-right clear region of the corresponding panel). This behaviour is consistent with the TNR results shown in \autoref{fig:MC-interference}. An ideal Recall map would display a dark strip at low mixing ratios of the target molecule; this is nearly the case for ozone when analysed for interference from water (second panel in the second row).

\subsubsection{MC for airless and CO$_2$-rich planets}
\label{uusec:MC-airless}

Following a similar analysis to that described in \autoref{uusec:BC-airless}, but now applied to the classification of individual molecules (CH$_4$, O$_3$, and H$_2$O), we evaluated the performance of MC in distinguishing two background scenarios: airless planets and CO$_2$-dominated atmospheres without other molecules. For airless scenarios, TNR values exceed 0.8 even in the most contamination-affected classifications at an SNR of 1, although significant improvement is only observed at an SNR of 5, where TNR values approach 0.9 for all molecules, even in the most challenging cases. In the case of CO$_2$-rich atmospheres, the classifier also exhibits robust performance. At an SNR of 1, TNR values remain above 0.7 in the worst-case scenario, with CH$_4$ and H$_2$O achieving values above 0.75, while O$_3$ drops to 0.7—likely due to the overlap between its main absorption peak and one of the CO$_2$ spectral bands. However, at an SNR of 3, TNR values exceed 0.8 for all molecules, confirming that the model maintains reliable classification performance under stellar contamination and in CO$_2$-dominated atmospheric conditions.

\subsection{Specialised Classifier}
\label{sec:SC}

To address the multilabel classification problem from another perspective, three RF binary classifiers were trained, each specialised in detecting a specific molecule: CH$_4$, H$_2$O, or O$_3$. The classification of planets according to the presence of all molecules using these individual specialised algorithms will be referred to as the \emph{Specialised Classifier} (SC).

SC is considerably more scalable than previous strategies. On one hand, if we want to add new molecules and/or bioindicators to the classification effort, it only requires training another specialised RF and adding it to the existing set of algorithms. On the other hand, it demands a lower number and complexity of training spectra, as explained below.

Unlike the approach in \autoref{sec:MC}, each RF was trained using a dataset containing spectra of airless planets, planets without bioindicators (i.e., only filler gases), and planets containing filler gases plus the molecule of interest. The numbers of training and test spectra for each molecule, depending on the SNR value, are presented in \autoref{tbl:data-SC}.

\begin{tiny}  
\begin{table*}
\caption{\label{tbl:data-SC}Training dataset distribution for detecting biosignatures/bioindicators. The distribution of spectra across SNR values is identical for each molecule. Labels are defined as {\tt mol} for spectra in which the molecule is present and {\tt $\neg\;$mol} for spectra in which the molecule is absent (i.e., only filler gases and airless). Each pair of columns shows the number of spectra at different SNRs, and the final column summarises the total number of spectra per molecule.}
\centering
\begin{tabular}{cccccccccccccc|c}
\hline
&\textbf{SNR}&\multicolumn{2}{c}{\textbf{1}}&\multicolumn{2}{c}{\textbf{3}}&\multicolumn{2}{c}{\textbf{6}}&\multicolumn{2}{c}{\textbf{10}}&\multicolumn{2}{c}{\textbf{0}}&\textbf{Total}\\
\hline
&Label&\textbf{\tt $\neg\;$mol}&\textbf{\tt mol}&\textbf{\tt $\neg\;$mol}&\textbf{\tt mol}&\textbf{\tt $\neg\;$mol}&\textbf{\tt mol}&\textbf{\tt $\neg\;$mol}&\textbf{\tt mol}&\textbf{\tt $\neg\;$mol}&\textbf{\tt mol}&\\
\hline
\textbf{Molecule} &&400000 & 240000 & 280000 & 240000 & 200000 & 120000 & 200000 & 120000  & 480000 & 400000 & 2680000 \\

\hline
\end{tabular}
\end{table*}
\end{tiny}

Each RF model consisted of 100 estimators that used the entropy criterion, had a maximum depth of 200, and a minimum samples leaf of 3. The classification thresholds for CH$_4$, O$_3$, and H$_2$O were set at 0.32, 0.24, and 0.37, respectively (for an explanation about the selection of the classification thresholds, see \autoref{sec:RF}).

To analyse the performance of the SC algorithm in classifying realistic spectra—namely, those containing a complex mixture of molecular species (including interfering ones)—we generated a test dataset of spectra using the same procedure described in \autoref{sec:MC}. Similarly, the same analyses and plots were generated for the SC as in the case of the MC. This ensures consistency in the comparative assessment of the models' performance.

\subsubsection{SC multilabel metrics}
\label{sec:SC-multilabel}

In \autoref{fig:SC-mulmetric}, we show the result of calculating the multilabel classification metrics for the SC, using a common autoencoder (upper panel) and a specialised autoencoder strategy (lower panel). The latter strategy employs a set of autoencoders that are trained on spectra containing only one molecule at a time. This contrasts with the common autoencoder, used for our BC and MC algorithms, which is trained on spectra containing all types of molecules. Although the specialised autoencoder strategy is more CPU intensive—since, when classifying a spectrum with respect to a given molecule, the corresponding specialised autoencoder is required—there is a trade-off between a higher classification overhead and lower CPU usage during training with smaller datasets.

When comparing the results of the SC with those of the MC (see \autoref{fig:MC-mulmetric}), we note that despite differences in training set complexity and scalability, the performance of both approaches is nearly identical for most metrics. This implies that training a set of specialised classifiers, with the advantage of easily adding new molecules as needed, yields results comparable to those of a more complex and CPU-intensive algorithm that handles all molecules simultaneously (the MC algorithm).

The most important differences between the common and specialised autoencoder strategies are observed in the Chemical Disequilibrium Recall (black continuous line and shaded area in both panels). We see that using specialised autoencoders significantly improves the algorithm's performance in correctly classifying planets with the considered biosignatures. Nevertheless, the SC model performs worse and is more sensitive to stellar contamination than the MC for this particular metric. Despite this drawback, it is remarkable that the SC algorithm is able to correctly classify most of the inhabited planets starting at SNR \(\approx 2\text{–}6\), which corresponds to fewer than a few tens of transits.

In a more general case with a realistic spectrum, the MC approach will show its advantages over the SC, as we will see in \autoref{sec:realistic}.

\begin{figure}
    \centering
    \includesvg[width=1\linewidth]{images/auto-SC-mulmetrics-v3.svg}
    \caption{Multilabel metrics for the SC algorithm as a function of SNR. The top panel shows results obtained with the common autoencoder (AE), while the bottom panel corresponds to the specialised AE. The most notable difference between these approaches is seen in the chemical disequilibrium recall metric, which improves significantly under the specialised AE strategy. For an explanation of the shaded regions, see \autoref{fig:interesting-gmetrix}. \label{fig:SC-mulmetric}}

\end{figure}

\subsubsection{SC molecule metrics}
\label{sec:SC-molmetrics}

In \autoref{fig:SC-molmetric}, we perform a similar comparison but evaluate the metrics for each molecule separately. This figure shows the results obtained with the common autoencoder strategy; the results for the specialised autoencoder strategy are not significantly different. The analogous results for the MC were presented in \autoref{fig:MC-molmetric}.

Again, the performance of both algorithms is comparable, at least for methane and ozone. In the case of water, however, the SC algorithm performs poorly compared to the MC. This finding implies that, when classifying molecules individually, the more scalable SC algorithm can be as effective as the MC algorithm for most molecules.

\begin{figure}
    \centering
    \includesvg[width=1\linewidth]{images/auto-SC-molmetrics.svg}
    \caption{Molecule-specific metrics from the SC using a common autoencoder. Each subplot shows the performance of different metrics (F1 Score, Recall, Precision, TNR) for each molecule (CH$_4$, O$_3$, H$_2$O) as the SNR increases. For an explanation of the shaded regions, see \autoref{fig:interesting-gmetrix}.}
    \label{fig:SC-molmetric}
\end{figure}

\subsubsection{SC minimum mixing ratio for detection}
\label{sec:SC-minmix}

In \autoref{fig:SC-minmix}, we show the minimum mixing ratio for detection achieved by the SC algorithm for each molecule. Here, we present the results obtained with the specialised autoencoder strategy, which outperforms the common autoencoder. The analogous figure for the MC algorithm is shown in \autoref{fig:MC-minmix}. As shown, the SC algorithm achieves slightly lower minimum mixing ratios for CH$_4$ and O$_3$ compared with the MC at lower SNRs.

\begin{figure}
    \centering
    \includesvg[width=1.07\linewidth]{images/auto-SC-minmix_v2.svg}
    \caption{Minimum mixing ratio required for the detection of each molecule (with Recall > 60\%) using SC classifiers and specialised autoencoders across different SNR values.}
    \label{fig:SC-minmix}
\end{figure}

\subsubsection{SC interference analysis}
\label{sec:SC-interference}

\begin{figure*}
    \centering
    \includegraphics[width=1\linewidth]{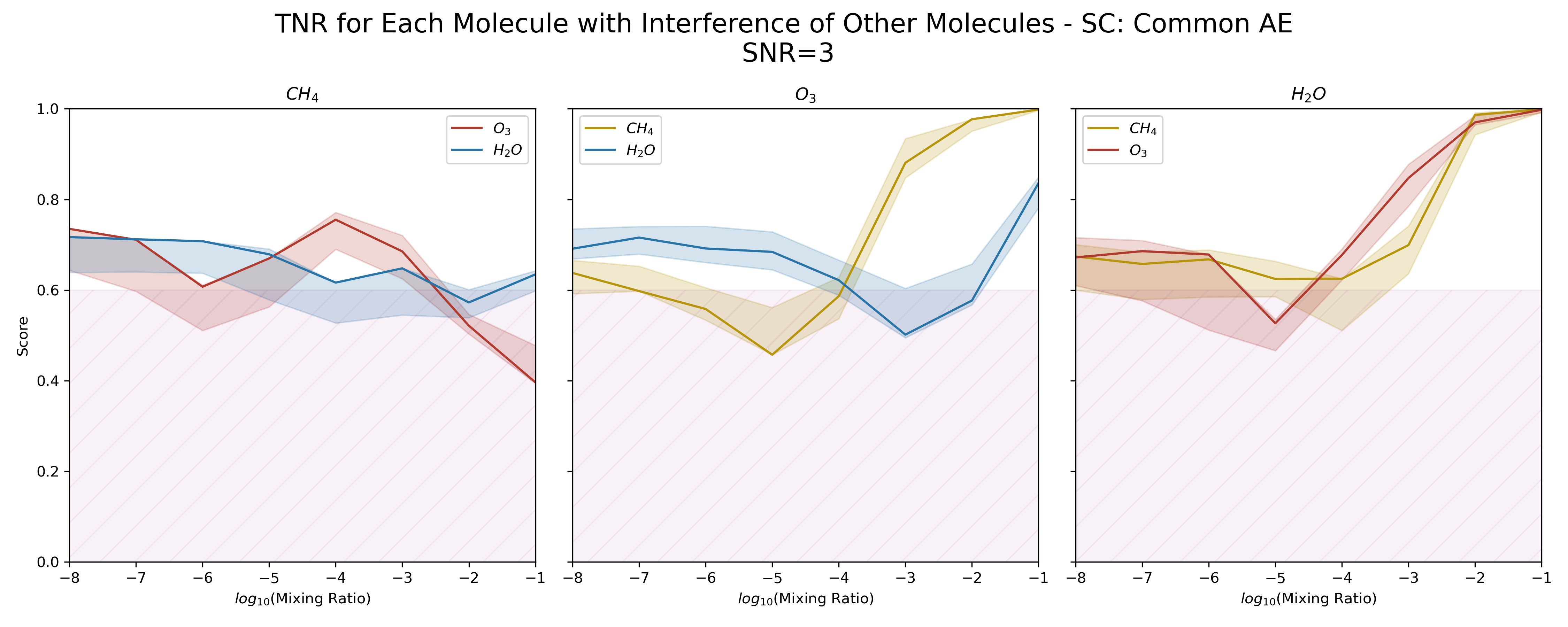}
    \caption{Same as \autoref{fig:MC-interference}, but for the case of SC algorithms with a common autoencoder.}
    \label{fig:SC-interference}
\end{figure*}

In \autoref{fig:SC-interference}, we analyse the effect that interfering molecules have on the detection of each target molecule for the SC algorithm. This analysis specifically addresses cases where the molecule of interest is absent and an interfering molecule may confuse the algorithm. A similar analysis was performed for the MC algorithm and is presented in \autoref{fig:MC-interference}. For details on the design of both figures, please refer to the explanation in \autoref{uusec:MC-interference}.

Compared to the MC case, the SC algorithm tends to be more conservative at correctly classifying the absence of target molecules—that is, the TNR values are consistently above the threshold for all tested interfering mixing ratios.

Interestingly, and in contrast to the MC case, large amounts of interference from other molecules improve the algorithm's performance in the detection of ozone and water (as evidenced by bumps at large mixing ratios in the middle and rightmost panels). This effect can be explained by considering that an algorithm specialised in detecting a given set of spectral signatures (for instance, those produced by ozone) will tend to recognise when significant spectral features of a different molecule are present, thus discounting those bands as belonging to the target molecule. It is as if the algorithm is "blinded" by the interfering molecule. This phenomenon is well illustrated by the differences observed in the interference on ozone detection produced by methane and water (middle panel). In particular, at a given mixing ratio, methane has a larger blinding effect on the ozone SC, likely because the spectral features of methane are more salient than those of water, at least in the studied spectral band. This is well illustrated in the spectrum shown in \autoref{fig:spec-dataset}.

In \autoref{fig:SC-RecallMaps}, we analyse the effect of interference on the Recall metric, specifically for the positive detection of each molecule. The corresponding figure for the MC algorithm is presented in \autoref{fig:MC-RecallMaps}.

\begin{figure*}
    \centering    
    \includesvg[width=0.9\linewidth]{images/auto-SC-RecallMap_v2.svg}
    \caption{Grid of Recall for SC with specialized autoencoders at SNR=3. Same as \autoref{fig:MC-RecallMaps}.}
    \label{fig:SC-RecallMaps}
\end{figure*}

As expected from the behaviour of the TNR observed in \autoref{fig:SC-interference}, the SC algorithms tend to respond negatively to the presence of interfering molecules (darker strips at high mixing ratios in the panels of the first column). However, this interference is less pronounced when the interfering molecule is water (panels of the second column).

The significant mutual interference between ozone and methane observed in the SC algorithm explains why the Chemical Disequilibrium Recall for this family of algorithms is lower than that of the MC (see \autoref{fig:SC-mulmetric}). In other words, in planets where both molecules are present, if one exists at a high mixing ratio, the SC algorithm tends to classify the planet as if the other molecule were absent, potentially discarding the most relevant astrobiological cases.

\subsubsection{SC for airless and CO$_2$-rich planets}
\label{uusec:SC-airless}

Applying the same analysis as in \autoref{uusec:MC-airless}, but now for SC classifiers, we find that for airless planets both autoencoder approaches perform well, though notable differences exist. The specialised autoencoder yields a TNR near 0.9 for all molecules at SNR = 1, whereas the common autoencoder produces more variable results. At SNR = 3, stellar contamination can reduce the TNR for O$_3$ and H$_2$O to about 0.8 in the worst cases; beyond this point, performance improves as SNR increases. For CH$_4$, the common autoencoder shows particularly strong performance, reaching TNR values above 0.9 at SNR = 5.

In CO$_2$-dominated atmospheres, CH$_4$ classification remains stable, with TNR around 0.7 at SNR = 1 for both autoencoder strategies and exceeding 0.8 at SNR = 5. O$_3$ and H$_2$O exhibit more strategy-dependent behaviour: at SNR = 1, the common autoencoder achieves a TNR of about 0.6 in the most contaminated cases, while the specialised approach maintains values above 0.65. Although TNR improves with increasing SNR for both strategies, the performance gap narrows from SNR = 3 onward.

\section{The SNR value and the number of transits}
\label{sec:SNRtransits}

Throughout the paper, we have used SNR to quantify the precision of the transmission spectra when training and testing the algorithms. In a real-life scenario, for example when performing observations using JWST, the noise level of an observed spectrum will be determined by the number of transits that can be measured during an observing program or survey campaign. In order to compare our results to those obtained in other works, we should estimate a relationship between the number of transits and the SNR of the resulting transmission spectra, at least for the scientific case we are studying here.

In \autoref{fig:SNR-Ntransits} we show the correspondence between the number of transits for TRAPPIST-1 e, hypothetically observed with the JWST NIRSpec PRISM, and the resulting SNR as computed with {\tt PandExo} \citep{batalhaPandExoCommunityTool2017}. To create this plot, we first calculate, using {\tt TauREx}, a theoretical spectrum for a TRAPPIST-1 e twin having an N$_2$ atmosphere with CO$_2$ at a mixing ratio of \(10^{-2}\). Then, we synthesise the observed transmission spectrum of the planet using the observing program and instrument parameters chosen by \cite{linDifferentiatingModernPrebiotic2021}. We repeat the calculation for a number of transits ranging from 1 to 100.

\begin{figure}
    \centering
    \includesvg[width=1\linewidth]{images/auto-SNR-Ntransit.svg}
    \caption{Characteristic SNR values (computed at 4.28 $\mu$m) of the transmission spectra of a TRAPPIST-1 e twin with an N$_2$-CO$_2$ atmosphere as a function of the number of transits observed with the JWST NIRSpec PRISM. Horizontal lines indicate the detection thresholds for H$_2$O (blue), CH$_4$ (yellow), and O$_3$ (red). The solid line represents the BC, while dashed lines correspond to the MC and dotted lines to the SC. This graph demonstrates that achieving an SNR above 10 would require an excessive number of transits, highlighting the challenges in detecting trace gases in exoplanetary atmospheres.}
    \label{fig:SNR-Ntransits}
\end{figure}

Since the noise of this instrument depends on wavelength, defining a global SNR is not trivial. For our analysis, we are interested in the spectral peaks that provide information about the presence of molecules. It is more reasonable to use the SNR of the signal at one of these peaks than at other parts of the spectrum (for instance, at the Rayleigh slope). For this reason, we use as a characteristic SNR the value corresponding to the CO$_2$ peak at 4.28 $\mu$m. This peak is sufficiently close to the maximum detectable wavelength of the instrument, where the noise is also higher. We acknowledge that, given the complexities of the instrumentation, the actual SNR value may vary under different conditions.

It is not hard to verify that the SNR value in \autoref{fig:SNR-Ntransits} grows as $\sqrt{N}$, where $N$ is the number of transits. This behaviour—although not immediately evident in the figure due to the semi-log scale—is expected, as an increase in the number of transits contributes to an increase in the number of data points measured at a given channel in the flux. The main consequence of this behaviour is that doubling the SNR requires almost four times as many transits. This represents one of the most significant instrumental challenges we face when searching for biosignatures.

\cite{linDifferentiatingModernPrebiotic2021} have calculated that, for detecting methane at 2.5$\sigma$ using standard retrieval procedures in a best-case scenario (mixing ratio greater than $10^{-5}$), a minimum of 10 transits (SNR > 3) is required. However, achieving similar precision in the retrieval of ozone requires as many as 200 transits, which, according to our estimation, corresponds to an SNR > 24.

If we are able to develop, as proposed in this paper, a methodology capable of selecting interesting targets with transmission spectra having SNR < 3 for methane and SNR < 24 for ozone, a significant improvement in our capabilities for searching for biosignatures/bioindicators could be achieved. These are precisely the SNR ranges at which we have conducted our numerical experiments.

To illustrate the capabilities of our method, we have included in \autoref{fig:SNR-Ntransits} the SNR required to confidently classify planets at different biosignature mixing ratios using the BC, MC and SC algorithms.

We find that, for classifying a planet as interesting when it contains ozone at levels as low as \(10^{-7}\) — nearly one order of magnitude lower than present Earth levels, and the minimum detectable mixing ratio by our algorithms — the MC algorithm requires 30–40 transits, the BC algorithm 20–30, while the SC requires only 10–20 transits.

Although a direct comparison with other detection strategies is challenging, it is noteworthy that our best algorithm requires almost 6–8 times fewer transits for the reliable classification of most interesting planets than do standard retrieval procedures (see, e.g., \citealt{linDifferentiatingModernPrebiotic2021}).

In the case of methane, the number of transits required for reliable classification using our methods (i.e., 10–20 transits) is similar to that achieved by some standard retrieval procedures. However, the minimum mixing ratio detectable by our algorithms is typically ten times lower. While this represents an extreme case, it illustrates the power of our approach.

In summary, SNR values between 3 and 5—where our algorithms reach acceptable performance—correspond to a number of transits between 10 and 30, which is feasible for a regular JWST observing programme. With a larger number of transits (30–50), our methods can classify planets as interesting at lower methane mixing ratios than those required by standard retrieval procedures using the same number of transits. The case of ozone is even more compelling; the MC and SC algorithms may tag a planet as interesting using up to six times fewer transits than those required for conventional retrieval procedures.

\section{Classifying realistic spectra}
\label{sec:realistic}

In order to test our algorithms with realistic cases, we use the high-resolution spectra of Earth at different geologic eras calculated by \cite{kalteneggerHighresolutionTransmissionSpectra2020}. For these spectra, we synthesise observations for JWST-NIRSpec using {\tt Pandexo}. The synthesis followed the same procedure we apply to calculate the relationship between SNR and the number of transits in \autoref{sec:SNRtransits}.

It is important to stress that the realistic spectra used as inputs for these experiments were originally calculated for a planet with a radius and surface gravity identical to Earth's. To apply these spectra to the slightly larger TRAPPIST-1 e, we simply rescale the line depths in the original spectrum (expressed in kilometres) to the radius of our planet. Of course, the photochemical equilibrium composition of Earth's atmosphere and that of TRAPPIST-1 e—which orbits an M-dwarf star and has different gravity and surface pressure—will be different. Nonetheless, since our purpose here is to test the performance of our algorithm when dealing with an actual planetary spectrum, assuming similar atmospheres does not violate any physical law.

For a given number of transits (i.e. a given characteristic SNR), we generate a sample of synthetic observations. To this end, we add stellar contamination to the theoretical signal using the procedure described in \autoref{sec:gen-data}. For each contaminated spectrum, we then synthesise 1,000 observed spectra, yielding a total of 10,000 test signals for each SNR (number of transits) value.

Once we have the test samples, we use the algorithms trained in previous sections (see \autoref{sec:binaryclass}, \autoref{sec:MC} and \autoref{sec:SC}) to classify the planet as interesting or not interesting, or as having or not having specific molecules, respectively.

\begin{figure}
    \centering
    \includegraphics[width=1\linewidth]{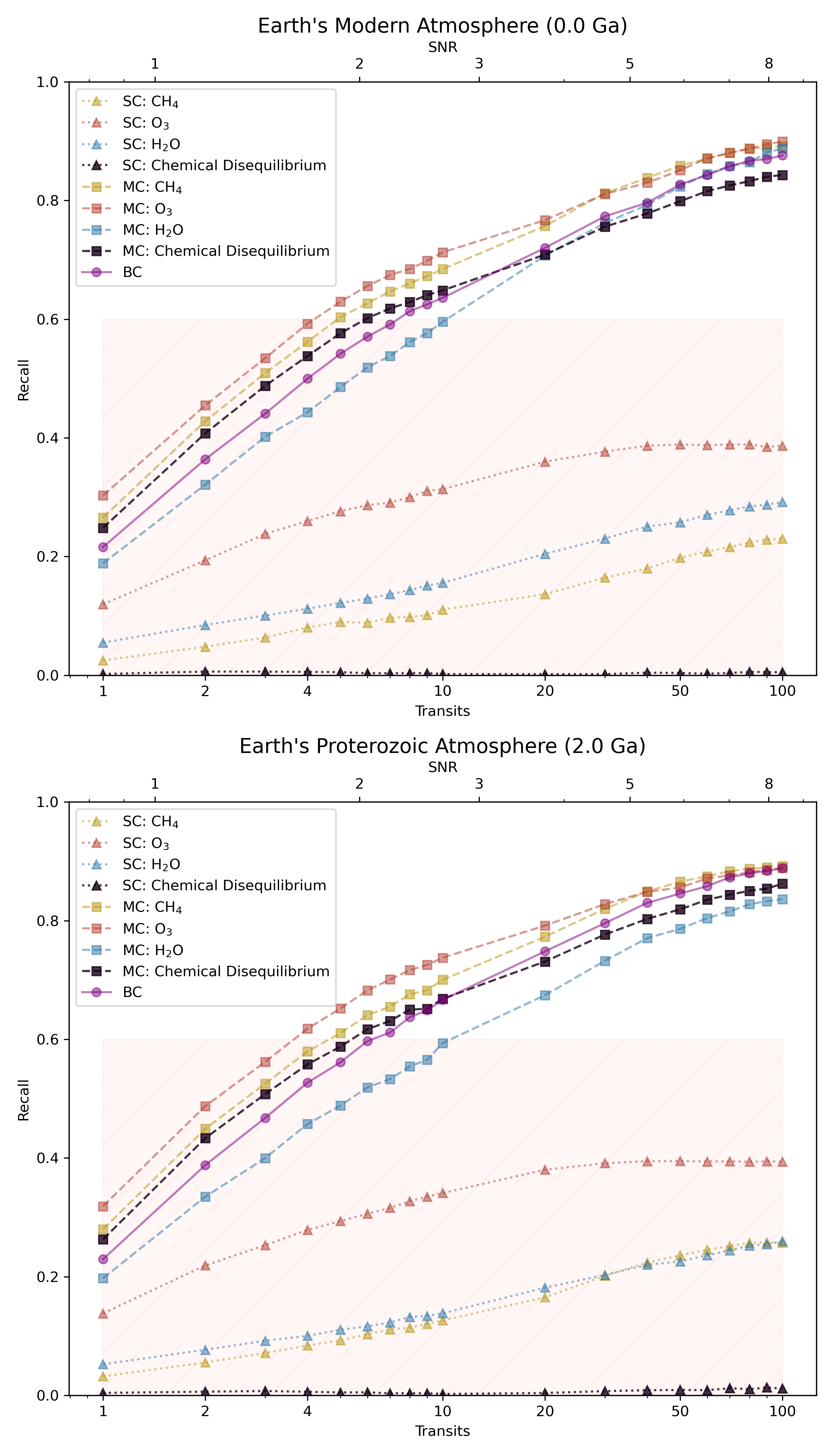}
    \caption{Recall as a function of the number of transits calculated for a set of 10,000 synthetic transmission spectra of a TRAPPIST-1e twin, covering a wide range of stellar contamination scenarios. These spectra were generated starting from high-resolution spectra of Earth's modern atmosphere (upper panel) and Earth's Proterozoic atmosphere (lower panel). Dotted lines with squares and dashed lines with triangles represent the performance of MC and SC classifiers for CH$_4$, O$_3$, H$_2$O, and chemical disequilibrium (simultaneous CH$_4$+O$_3$). The BC algorithm is shown as a continuous (purple) line.}
    \label{fig:realistic-test}
\end{figure}

In \autoref{fig:realistic-test}, we show the Recall metric obtained for a set of synthetic spectra of a TRAPPIST-1 e planet with a modern Earth's atmosphere (upper panel) and a Proterozoic Earth's atmosphere (lower panel).

In previous analysis we showed the value of the metrics obtained for different degrees of contamination considered separately. This gave us the shaded bands in \autoref{fig:TNR-water}-\autoref{fig:SC-interference}. For this analysis, and for the sake of simplifying the analysis of the results, we obtain the recall of the full set, independently of the level of contamination.

We observe that when dealing with more realistic spectra, namely an atmosphere with a more complex physical structure and a greater number of molecules, the binary (BC) and multilabel classifiers (MC) perform well even if the number of transits is as low as 4-10.

This result suggests that small-to-medium-scale observing campaigns, combined with classification-oriented machine-learning algorithms rather than full atmospheric retrievals, can effectively identify planets warranting detailed follow-up observations. 

Our results demonstrate that even an algorithm that was trained only with synthetic and simple spectra, can correctly classify more than 60\% of the planets with biosignatures/bioindicators in a survey sample. We speculate, that with a more complex training set and probably more competent methods, the performance of machine learning models using our methodology could be significantly enhanced.

The specialized classifier (SC, dotted lines) performs worse when dealing with realistic spectra, than the binary (BC, continuous line) and multilabel (MC, dashed lines) classifier. 

This outcome arises because our specialised algorithm is overfitted to classify less complex spectra with a simplified noise structure,. However, the realistic spectra we are using for this test has many molecular species and a wavelength-dependent instrumental noise. It is interesting, however, to notice how well the BC and MC algorithms performs when dealing with such complex spectra, even although they were also trained with less realistic signals.

As expected, the values of the SNR where the BC and MC algorithms reach the threshold of 60\% spectra correctly classified, are consistent with the minimum mixing ratio for detection (see \autoref{fig:interesting-minmix} and \autoref{fig:MC-minmix}) for the levels of methane and ozone in the modeled Earth.

\section{Discussion}
\label{sec:discussion}

The validity of the results presented in this work may depend on the specific properties of the planet, its atmosphere, and the star selected for our numerical experiments. In the following, we analyse the sensitivity of our results to these variable factors.

If we disregard the effect of atmospheric refraction (see below), changing the radius of the star or the planet simply alters the signal strength and noise, and thus the SNR value. Pressure, which essentially determines the atmospheric scale height, has a similar effect. Since our experiments span a broad range of SNR values, any change in these parameters will lie within one of the synthetic noisy spectra used for training and testing. In other words, a variation in SNR is equivalent to a variation in either the planet-to-star radius ratio or the atmospheric pressure. Although these factors influence equilibrium compositions in realistic atmospheric models, the simplified model atmospheres employed in this study do not explicitly link atmospheric chemistry to these parameters. We emphasise that our aim is to demonstrate the feasibility of this approach, rather than to design robust algorithms for an actual observing campaign.

Other studies (e.g. \citealt{betremieuxIMPACTATMOSPHERICREFRACTION2014}) have shown that atmospheric refraction limits how deeply an exoplanet’s atmosphere can be probed, and that this limitation depends on the host star’s radius. For M-dwarf stars—which appear smaller from the perspective of a habitable-zone planet (see Figure 2 in \citealt{betremieuxIMPACTATMOSPHERICREFRACTION2014})—the refractive effect is minimal, and stellar light can reach very deep atmospheric layers, sometimes down to the surface. Conversely, for K-dwarf stars, the larger angular size results in a higher critical altitude. For instance, in the case of a K0 star, stellar light may only penetrate to about 8–14\,km, which, for present-day Earth, lies above the troposphere where methane and water are most abundant.

Although our model does not explicitly include these refractive effects, the results of such studies allow us to affirm generally that the host star’s spectral type influences the accessible atmospheric depth—minimal for M-dwarfs but more significant for K-dwarfs, particularly if clouds are present.

Changing the star’s effective temperature (and thus its spectral energy distribution) or the planetary atmospheric temperature would not only affect the SNR but also modify the transmission spectra in a non-trivial manner, potentially impacting algorithm performance. To test the effect of varying stellar spectra, we conducted additional numerical experiments with different stellar effective temperatures in the M- and K-dwarf ranges. We find that altering the stellar spectrum does not significantly affect classification results.

On the other hand, the effect of atmospheric temperature was already implicit in our experiments. As described, for instance, in \autoref{sec:binaryclass}, our training and test sets include transmission spectra at various atmospheric temperatures (from 200\,K to 400\,K). We observe that at lower temperatures, algorithm performance improves. For example, at 200\,K, the minimum mixing ratio required for detection can be reduced by an order of magnitude at very low SNR values for several molecules. This temperature dependence arises because molecular cross-sections—especially that of CO$_2$—change with temperature, affecting the algorithm’s ability to recognise molecular signatures.

In summary, our models are robust against variations in stellar temperature and radius, at least for M-dwarfs. The case of K-dwarfs should be examined independently to account for atmospheric refraction. We anticipate that this effect would merely shift the mixing ratio thresholds obtained here without altering the main conclusions of this work. Our experiments indicate that planetary atmospheric temperature can  affect the accuracy of biosignature/bioindicator classification.

Since the performance of the SC algorithms depends on the noise structure, one might ask why we did not train our algorithms using the specific noise profile of NIRSpec. We reiterate that our objective was to assess the feasibility of using machine learning to classify planets with biosignatures/bioindicators at low SNR, rather than optimising for a particular noise structure. Focusing on a specific type of noise could render our results noise-dependent and undermine the generality of our conclusions. Nevertheless, further investigation of how the noise structure affects classification algorithm performance is warranted.


\section{Summary and conclusions}
\label{sec:conclusion}

In this work, we explored the use of supervised machine learning algorithms to classify low-SNR exoplanet transmission spectra based on their potential for containing biosignatures/bioindicators. We employed Random Forest (RF) to design and train three types of models: a binary classifier (BC) that labels a planet as interesting or not for follow-up observations; a multilabel classifier (MC) that assigns labels to an observed spectrum according to the presence of certain molecular species; and specialised classifiers (SC) that operate similarly to MC by assembling multiple binary RF classifiers focused on particular molecules.

For training and testing, we generated on the order of \(10^{7}\) synthetic noisy spectra, including stellar contamination, corresponding to a planet with physical characteristics similar to TRAPPIST-1 e, but covering a wide diversity of compositions and three atmospheric temperatures.

Generating and handling such a large set of spectra is challenging. For this purpose, we developed a package, {\tt MultiREx}\footnote{\url{http://pypi.org/project/multirex}}, which serves as a wrapper for {\tt TauREx} \citep{al-refaieTauRExFastDynamic2021} and is specifically designed to generate, represent, and manipulate large sets of synthetic noisy spectra for machine learning applications.

We tested our models by measuring various binary and multilabel standard metrics (Recall, TNR, F1 Score, etc.), which we interpret in the context of biosignature/bioindicator searches as time-saving metrics (minimising the time spent on following up uninteresting planets), wasting metrics (ensuring inhabited planets are not erroneously excluded), discovery metrics (maximising the opportunity for actual discovery), among others.

Although the BC algorithm lacks the precision for detailed molecular classification, it achieves the highest values for the discovery metric (F1 Score). With observed spectra having SNR values as low as 1 or 2—which, according to our estimations, can be achieved by observing 1–5 transits with the JWST NIRSpec PRISM—the BC model labels most of the inhabited planets in a survey sample as interesting, even when they exhibit a wide range of biosignature/bioindicator mixing ratios. Moreover, the algorithm performs exceptionally well (TNR > 80\%) at tagging planets without biosignatures/bioindicators as not interesting, thereby avoiding wasted follow-up observations.

Our results indicate that by surveying potentially habitable exoplanets with transmission spectra obtained from only 4 to 10 transits, we can identify most inhabited exoplanets for follow-up, time-intensive observations—provided that their atmospheres have integrated mixing ratios of CH$_4$ and O$_3$ similar to those of present and prebiotic Earth (\(\mathcal{O}(10^{-6})\) and \(\mathcal{O}(10^{-5})\), respectively; \citealt{lustig-yaegerEarthTransitingExoplanet2023, linDifferentiatingModernPrebiotic2021}). This number of transits is significantly lower than that required by standard and non-standard retrieval procedures for detecting these molecules.

It must be emphasised that the methodology proposed here is not intended to replace more detailed retrieval methods or to directly detect biosignatures/bioindicators. Rather, it aims to enhance the efficiency of observing time allocation for valuable resources such as JWST by identifying interesting targets for follow-up observations—an important goal in modern astronomy.

For future research, it would be valuable to explore the integration of additional machine learning methods and to increase the diversity of the synthetic spectra used for training. Furthermore, incorporating other potential biosignatures or bioindicators, along with a broader range of atmospheric conditions, could enhance the models' generalisability. These advances will contribute to more accurate and efficient detection of biosignatures/bioindicators, thereby supporting ongoing efforts in the search for life beyond our Solar System.
\section*{Acknowledgments}

We sincerely acknowledge the referee for their insightful comments and valuable suggestions, especially for proposing additional aspects to evaluate and train our models, which significantly enhance their robustness.

\section*{Data availability and reproducibility}
\label{sec:data-availability}

All data required to replicate the results and generate the figures presented in this work, including the {\tt Jupyter} notebooks used to train and test the  models, are available in the {\tt GitHub} public repository of the package {\tt MultiREx} at \url{https://github.com/D4san/MultiREx-public}.

\bibliographystyle{mnras}
\bibliography{references} 

\begin{thebibliography}{}
\makeatletter
\relax
\def\mn@urlcharsother{\let\do\@makeother \do\$\do\&\do\#\do\^\do\_\do\%\do\~}
\def\mn@doi{\begingroup\mn@urlcharsother \@ifnextchar [ {\mn@doi@}
  {\mn@doi@[]}}
\def\mn@doi@[#1]#2{\def\@tempa{#1}\ifx\@tempa\@empty \href
  {http://dx.doi.org/#2} {doi:#2}\else \href {http://dx.doi.org/#2} {#1}\fi
  \endgroup}
\def\mn@eprint#1#2{\mn@eprint@#1:#2::\@nil}
\def\mn@eprint@arXiv#1{\href {http://arxiv.org/abs/#1} {{\tt arXiv:#1}}}
\def\mn@eprint@dblp#1{\href {http://dblp.uni-trier.de/rec/bibtex/#1.xml}
  {dblp:#1}}
\def\mn@eprint@#1:#2:#3:#4\@nil{\def\@tempa {#1}\def\@tempb {#2}\def\@tempc
  {#3}\ifx \@tempc \@empty \let \@tempc \@tempb \let \@tempb \@tempa \fi \ifx
  \@tempb \@empty \def\@tempb {arXiv}\fi \@ifundefined
  {mn@eprint@\@tempb}{\@tempb:\@tempc}{\expandafter \expandafter \csname
  mn@eprint@\@tempb\endcsname \expandafter{\@tempc}}}

\bibitem[\protect\citeauthoryear{Abadi et~al.,}{Abadi
  et~al.}{2015}]{tensorflow2015-whitepaper}
Abadi M.,  et~al., 2015, {TensorFlow}: Large-Scale Machine Learning on
  Heterogeneous Systems, \url {https://www.tensorflow.org/}

\bibitem[\protect\citeauthoryear{Agol et~al.,}{Agol
  et~al.}{2021}]{agolRefiningTransittimingPhotometric2021}
Agol E.,  et~al., 2021, \mn@doi [The Planetary Science Journal]
  {10.3847/PSJ/abd022}, 2, 1

\bibitem[\protect\citeauthoryear{Airapetian et~al.,}{Airapetian
  et~al.}{2020}]{airapetianImpactSpaceWeather2020}
Airapetian V.~S.,  et~al., 2020, \mn@doi [International Journal of
  Astrobiology] {10.1017/S1473550419000132}, 19, 136

\bibitem[\protect\citeauthoryear{{Al-Refaie}, Changeat, Waldmann  \&
  Tinetti}{{Al-Refaie} et~al.}{2021}]{al-refaieTauRExFastDynamic2021}
{Al-Refaie} A.~F.,  Changeat Q.,  Waldmann I.~P.,   Tinetti G.,  2021, \mn@doi
  [The Astrophysical Journal] {10.3847/1538-4357/ac0252}, 917, 37

\bibitem[\protect\citeauthoryear{Ard{\'e}vol~Mart{\'i}nez, Min, Kamp  \&
  Palmer}{Ard{\'e}vol~Mart{\'i}nez
  et~al.}{2022}]{ardevolmartinezConvolutionalNeuralNetworks2022}
Ard{\'e}vol~Mart{\'i}nez F.,  Min M.,  Kamp I.,   Palmer P.~I.,  2022, \mn@doi
  [Astronomy \& Astrophysics] {10.1051/0004-6361/202142976}, 662, A108

\bibitem[\protect\citeauthoryear{Ard{\'e}vol~Mart{\'i}nez, Min, Huppenkothen,
  Kamp  \& Palmer}{Ard{\'e}vol~Mart{\'i}nez
  et~al.}{2024}]{ardevolmartinezFlopPITyEnablingSelfconsistent2024}
Ard{\'e}vol~Mart{\'i}nez F.,  Min M.,  Huppenkothen D.,  Kamp I.,   Palmer
  P.~I.,  2024, \mn@doi [Astronomy \& Astrophysics]
  {10.1051/0004-6361/202348367}, 681, L14

\bibitem[\protect\citeauthoryear{Barstow \& Irwin}{Barstow \&
  Irwin}{2016}]{barstowHabitableWorldsJWST2016}
Barstow J.~K.,  Irwin P. G.~J.,  2016, \mn@doi [Monthly Notices of the Royal
  Astronomical Society: Letters] {10.1093/mnrasl/slw109}, 461, L92

\bibitem[\protect\citeauthoryear{Batalha et~al.,}{Batalha
  et~al.}{2017}]{batalhaPandExoCommunityTool2017}
Batalha N.~E.,  et~al., 2017, \mn@doi [Publications of the Astronomical Society
  of the Pacific] {10.1088/1538-3873/aa65b0}, 129, 064501

\bibitem[\protect\citeauthoryear{Benneke et~al.,}{Benneke
  et~al.}{2024}]{bennekeJWSTRevealsCH$_4$2024}
Benneke B.,  et~al., 2024, {{JWST Reveals CH}}\$\_4\$, {{CO}}\$\_2\$, and
  {{H}}\$\_2\${{O}} in a {{Metal-rich Miscible Atmosphere}} on a
  {{Two-Earth-Radius Exoplanet}} (\mn@eprint {arXiv} {2403.03325})

\bibitem[\protect\citeauthoryear{Berahmand, Daneshfar, Salehi, Li  \&
  Xu}{Berahmand et~al.}{2024}]{berahmandAutoencodersTheirApplications2024}
Berahmand K.,  Daneshfar F.,  Salehi E.~S.,  Li Y.,   Xu Y.,  2024, \mn@doi
  [Artificial Intelligence Review] {10.1007/s10462-023-10662-6}, 57, 28

\bibitem[\protect\citeauthoryear{B{\'e}tr{\'e}mieux \&
  Kaltenegger}{B{\'e}tr{\'e}mieux \&
  Kaltenegger}{2014}]{betremieuxIMPACTATMOSPHERICREFRACTION2014}
B{\'e}tr{\'e}mieux Y.,  Kaltenegger L.,  2014, \mn@doi [The Astrophysical
  Journal] {10.1088/0004-637X/791/1/7}, 791, 7

\bibitem[\protect\citeauthoryear{Bourrier et~al.,}{Bourrier
  et~al.}{2017}]{bourrierTemporalEvolutionHighenergy2017}
Bourrier V.,  et~al., 2017, \mn@doi [The Astronomical Journal]
  {10.3847/1538-3881/aa859c}, 154, 121

\bibitem[\protect\citeauthoryear{Breiman}{Breiman}{2001}]{breimanRandomForests2001}
Breiman L.,  2001, \mn@doi [Machine Learning] {10.1023/A:1010933404324}, 45, 5

\bibitem[\protect\citeauthoryear{Burgasser \& Mamajek}{Burgasser \&
  Mamajek}{2017}]{burgasserAgeTRAPPIST1System2017}
Burgasser A.~J.,  Mamajek E.~E.,  2017, \mn@doi [The Astrophysical Journal]
  {10.3847/1538-4357/aa7fea}, 845, 110

\bibitem[\protect\citeauthoryear{Cadieux et~al.,}{Cadieux
  et~al.}{2024}]{cadieuxTransmissionSpectroscopyHabitable2024}
Cadieux C.,  et~al., 2024, Transmission {{Spectroscopy}} of the {{Habitable
  Zone Exoplanet LHS}} 1140 b with {{JWST}}/{{NIRISS}} (\mn@eprint {arXiv}
  {2406.15136})

\bibitem[\protect\citeauthoryear{Chollet}{Chollet}{2015}]{chollet2015keras}
Chollet F.,  2015, Keras, \url{https://github.com/fchollet/keras}

\bibitem[\protect\citeauthoryear{Chubb et~al.,}{Chubb
  et~al.}{2021}]{chubbExoMolOPDatabaseCross2021}
Chubb K.~L.,  et~al., 2021, \mn@doi [Astronomy \& Astrophysics]
  {10.1051/0004-6361/202038350}, 646, A21

\bibitem[\protect\citeauthoryear{Cobb et~al.,}{Cobb
  et~al.}{2019}]{cobbEnsembleBayesianNeural2019}
Cobb A.~D.,  et~al., 2019, \mn@doi [The Astronomical Journal]
  {10.3847/1538-3881/ab2390}, 158, 33

\bibitem[\protect\citeauthoryear{Des~Marais et~al.,}{Des~Marais
  et~al.}{2008}]{desmaraisNASAAstrobiologyRoadmap2008}
Des~Marais D.~J.,  et~al., 2008, \mn@doi [Astrobiology]
  {10.1089/ast.2008.0819}, 8, 715

\bibitem[\protect\citeauthoryear{{Domagal-Goldman}, Segura, Claire, Robinson
  \& Meadows}{{Domagal-Goldman}
  et~al.}{2014}]{domagal-goldmanABIOTICOZONEOXYGEN2014}
{Domagal-Goldman} S.~D.,  Segura A.,  Claire M.~W.,  Robinson T.~D.,   Meadows
  V.~S.,  2014, \mn@doi [The Astrophysical Journal]
  {10.1088/0004-637X/792/2/90}, 792, 90

\bibitem[\protect\citeauthoryear{Doshi, Cowan  \& Huang}{Doshi
  et~al.}{2022}]{doshiStratosphericCloudsNot2022}
Doshi D.,  Cowan N.~B.,   Huang Y.,  2022, \mn@doi [Monthly Notices of the
  Royal Astronomical Society] {10.1093/mnras/stac1869}, 515, 1982

\bibitem[\protect\citeauthoryear{Fauchez et~al.,}{Fauchez
  et~al.}{2020}]{fauchezSensitiveProbingExoplanetary2020}
Fauchez T.~J.,  et~al., 2020, \mn@doi [Nature Astronomy]
  {10.1038/s41550-019-0977-7}, 4, 372

\bibitem[\protect\citeauthoryear{Forestano, Matchev, Matcheva  \&
  Unlu}{Forestano et~al.}{2023}]{forestanoSearchingNovelChemistry2023}
Forestano R.~T.,  Matchev K.~T.,  Matcheva K.,   Unlu E.~B.,  2023, \mn@doi
  [The Astrophysical Journal] {10.3847/1538-4357/ad0047}, 958, 106

\bibitem[\protect\citeauthoryear{France et~al.,}{France
  et~al.}{2020}]{franceHighenergyRadiationEnvironment2020}
France K.,  et~al., 2020, \mn@doi [The Astronomical Journal]
  {10.3847/1538-3881/abb465}, 160, 237

\bibitem[\protect\citeauthoryear{Freedman, Marley  \& Lodders}{Freedman
  et~al.}{2008}]{freedmanLineMeanOpacities2008}
Freedman R.~S.,  Marley M.~S.,   Lodders K.,  2008, \mn@doi [The Astrophysical
  Journal Supplement Series] {10.1086/521793}, 174, 504

\bibitem[\protect\citeauthoryear{Freedman, {Lustig-Yaeger}, Fortney, Lupu,
  Marley  \& Lodders}{Freedman et~al.}{2014}]{freedmanGASEOUSMEANOPACITIES2014}
Freedman R.~S.,  {Lustig-Yaeger} J.,  Fortney J.~J.,  Lupu R.~E.,  Marley
  M.~S.,   Lodders K.,  2014, \mn@doi [The Astrophysical Journal Supplement
  Series] {10.1088/0067-0049/214/2/25}, 214, 25

\bibitem[\protect\citeauthoryear{Gao, Hu, Robinson, Li  \& Yung}{Gao
  et~al.}{2015}]{gaoSTABILITYCO2ATMOSPHERES2015}
Gao P.,  Hu R.,  Robinson T.~D.,  Li C.,   Yung Y.~L.,  2015, \mn@doi [The
  Astrophysical Journal] {10.1088/0004-637X/806/2/249}, 806, 249

\bibitem[\protect\citeauthoryear{Gebhard, Angerhausen, Konrad, Alei, Quanz  \&
  Sch{\"o}lkopf}{Gebhard
  et~al.}{2024}]{gebhardParameterizingPressureTemperature2024}
Gebhard T.~D.,  Angerhausen D.,  Konrad B.~S.,  Alei E.,  Quanz S.~P.,
  Sch{\"o}lkopf B.,  2024, \mn@doi [Astronomy \& Astrophysics]
  {10.1051/0004-6361/202346390}, 681, A3

\bibitem[\protect\citeauthoryear{Genest, Lafreni{\`e}re, Boucher,
  {Darveau-Bernier}, Doyon, Artigau  \& Cook}{Genest
  et~al.}{2022}]{genestEffectStellarActivity2022}
Genest F.,  Lafreni{\`e}re D.,  Boucher A.,  {Darveau-Bernier} A.,  Doyon R.,
  Artigau {\'E}.,   Cook N.,  2022, \mn@doi [The Astronomical Journal]
  {10.3847/1538-3881/ac5e38}, 163, 231

\bibitem[\protect\citeauthoryear{G{\'e}ron}{G{\'e}ron}{2023}]{geronHandsonMachineLearning2023}
G{\'e}ron A.,  2023, Hands-on Machine Learning with {{Scikit-Learn}},
  {{Keras}}, and {{TensorFlow}}: Concepts, Tools, and Techniques to Build
  Intelligent Systems, third edition edn.
O'Reilly, Beijing Boston Farnham Sebastopol Tokyo

\bibitem[\protect\citeauthoryear{Gillon}{Gillon}{2024}]{gillonTRAPPIST1ItsCompact2024}
Gillon M.,  2024, {{TRAPPIST-1}} and Its Compact System of Temperate Rocky
  Planets (\mn@eprint {arXiv} {2401.11815})

\bibitem[\protect\citeauthoryear{Gillon et~al.,}{Gillon
  et~al.}{2016}]{gillonTemperateEarthsizedPlanets2016}
Gillon M.,  et~al., 2016, \mn@doi [Nature] {10.1038/nature17448}, 533, 221

\bibitem[\protect\citeauthoryear{Gillon et~al.,}{Gillon
  et~al.}{2020}]{gillonTRAPPIST1JWSTCommunity2020}
Gillon M.,  et~al., 2020, \mn@doi [Bulletin of the AAS]
  {10.3847/25c2cfeb.afbf0205}, 52

\bibitem[\protect\citeauthoryear{Grenfell}{Grenfell}{2018}]{grenfellAtmosphericBiosignatures2018}
Grenfell J.~L.,  2018, in Deeg H.~J.,  Belmonte J.~A.,  eds, , Handbook of
  {{Exoplanets}}.
Springer International Publishing, Cham, pp 1--14,
  \mn@doi{10.1007/978-3-319-30648-3_68-1}

\bibitem[\protect\citeauthoryear{Grenfell, Gebauer, V.~Paris, Godolt  \&
  Rauer}{Grenfell et~al.}{2014}]{grenfellSensitivityBiosignaturesEarthlike2014}
Grenfell J.,  Gebauer S.,  V.~Paris P.,  Godolt M.,   Rauer H.,  2014, \mn@doi
  [Planetary and Space Science] {10.1016/j.pss.2013.10.006}, 98, 66

\bibitem[\protect\citeauthoryear{{Guzm{\'a}n-Marmolejo}, Segura  \&
  {Escobar-Briones}}{{Guzm{\'a}n-Marmolejo}
  et~al.}{2013}]{guzman-marmolejoAbioticProductionMethane2013}
{Guzm{\'a}n-Marmolejo} A.,  Segura A.,   {Escobar-Briones} E.,  2013, \mn@doi
  [Astrobiology] {10.1089/ast.2012.0817}, 13, 550

\bibitem[\protect\citeauthoryear{{Guzm{\'a}n-Mesa} et~al.,}{{Guzm{\'a}n-Mesa}
  et~al.}{2020}]{guzman-mesaInformationContentJWST2020}
{Guzm{\'a}n-Mesa} A.,  et~al., 2020, \mn@doi [The Astronomical Journal]
  {10.3847/1538-3881/ab9176}, 160, 15

\bibitem[\protect\citeauthoryear{Harman, Schwieterman, Schottelkotte  \&
  Kasting}{Harman et~al.}{2015}]{harmanABIOTICO2LEVELS2015}
Harman C.~E.,  Schwieterman E.~W.,  Schottelkotte J.~C.,   Kasting J.~F.,
  2015, \mn@doi [The Astrophysical Journal] {10.1088/0004-637X/812/2/137}, 812,
  137

\bibitem[\protect\citeauthoryear{Hayes et~al.,}{Hayes
  et~al.}{2020}]{hayesOptimizingExoplanetAtmosphere2020}
Hayes J. J.~C.,  et~al., 2020, \mn@doi [Monthly Notices of the Royal
  Astronomical Society] {10.1093/mnras/staa978}, 494, 4492

\bibitem[\protect\citeauthoryear{Himes et~al.,}{Himes
  et~al.}{2022}]{himesAccurateMachinelearningAtmospheric2022}
Himes M.~D.,  et~al., 2022, \mn@doi [The Planetary Science Journal]
  {10.3847/PSJ/abe3fd}, 3, 91

\bibitem[\protect\citeauthoryear{Huang~(黄硕) \& Ormel}{Huang~(黄硕) \&
  Ormel}{2022}]{huangHuangShuoDynamicsTRAPPIST1System2022}
Huang~(黄硕) S.,  Ormel C.~W.,  2022, \mn@doi [Monthly Notices of the Royal
  Astronomical Society] {10.1093/mnras/stac288}, 511, 3814

\bibitem[\protect\citeauthoryear{Husser, {Wende-von Berg}, Dreizler, Homeier,
  Reiners, Barman  \& Hauschildt}{Husser
  et~al.}{2013}]{husserNewExtensiveLibrary2013}
Husser T.-O.,  {Wende-von Berg} S.,  Dreizler S.,  Homeier D.,  Reiners A.,
  Barman T.,   Hauschildt P.~H.,  2013, \mn@doi [Astronomy \& Astrophysics]
  {10.1051/0004-6361/201219058}, 553, A6

\bibitem[\protect\citeauthoryear{Iyer \& Line}{Iyer \&
  Line}{2020}]{iyerInfluenceStellarContamination2020}
Iyer A.~R.,  Line M.~R.,  2020, \mn@doi [The Astrophysical Journal]
  {10.3847/1538-4357/ab612e}, 889, 78

\bibitem[\protect\citeauthoryear{Kaltenegger, Traub  \& Jucks}{Kaltenegger
  et~al.}{2007}]{kalteneggerSpectralEvolutionEarthlike2007}
Kaltenegger L.,  Traub W.~A.,   Jucks K.~W.,  2007, \mn@doi [The Astrophysical
  Journal] {10.1086/510996}, 658, 598

\bibitem[\protect\citeauthoryear{Kaltenegger, Lin  \& Madden}{Kaltenegger
  et~al.}{2020}]{kalteneggerHighresolutionTransmissionSpectra2020}
Kaltenegger L.,  Lin Z.,   Madden J.,  2020, \mn@doi [The Astrophysical
  Journal] {10.3847/2041-8213/ab789f}, 892, L17

\bibitem[\protect\citeauthoryear{Kelleher, Mac~Namee  \& D'Arcy}{Kelleher
  et~al.}{2015}]{kelleherFundamentalsMachineLearning2015}
Kelleher J.~D.,  Mac~Namee B.,   D'Arcy A.,  2015, Fundamentals of Machine
  Learning for Predictive Data Analytics: Algorithms, Worked Examples, and Case
  Studies.
The MIT Press, Cambridge, Massachusetts

\bibitem[\protect\citeauthoryear{L{\'e}ger, Fontecave, Labeyrie, Samuel,
  Demangeon  \& Valencia}{L{\'e}ger
  et~al.}{2011}]{legerPresenceOxygenExoplanet2011}
L{\'e}ger A.,  Fontecave M.,  Labeyrie A.,  Samuel B.,  Demangeon O.,
  Valencia D.,  2011, \mn@doi [Astrobiology] {10.1089/ast.2010.0516}, 11, 335

\bibitem[\protect\citeauthoryear{Lim et~al.,}{Lim
  et~al.}{2023}]{limAtmosphericReconnaissanceTRAPPIST12023}
Lim O.,  et~al., 2023, \mn@doi [The Astrophysical Journal Letters]
  {10.3847/2041-8213/acf7c4}, 955, L22

\bibitem[\protect\citeauthoryear{Lin \& Kaltenegger}{Lin \&
  Kaltenegger}{2022}]{linHighresolutionSpectralModels2022}
Lin Z.,  Kaltenegger L.,  2022, \mn@doi [Monthly Notices of the Royal
  Astronomical Society] {10.1093/mnras/stac2536}, 516, 3167

\bibitem[\protect\citeauthoryear{Lin, MacDonald, Kaltenegger  \& Wilson}{Lin
  et~al.}{2021}]{linDifferentiatingModernPrebiotic2021}
Lin Z.,  MacDonald R.~J.,  Kaltenegger L.,   Wilson D.~J.,  2021, \mn@doi
  [Monthly Notices of the Royal Astronomical Society] {10.1093/mnras/stab1486},
  505, 3562

\bibitem[\protect\citeauthoryear{Lueber, Kitzmann, Fisher, Bowler, Burgasser,
  Marley  \& Heng}{Lueber et~al.}{2023}]{lueberIntercomparisonBrownDwarf2023}
Lueber A.,  Kitzmann D.,  Fisher C.~E.,  Bowler B.~P.,  Burgasser A.~J.,
  Marley M.,   Heng K.,  2023, \mn@doi [The Astrophysical Journal]
  {10.3847/1538-4357/ace530}, 954, 22

\bibitem[\protect\citeauthoryear{Luger \& Barnes}{Luger \&
  Barnes}{2015}]{lugerExtremeWaterLoss2015}
Luger R.,  Barnes R.,  2015, \mn@doi [Astrobiology] {10.1089/ast.2014.1231},
  15, 119

\bibitem[\protect\citeauthoryear{Lupu et~al.,}{Lupu
  et~al.}{2014}]{lupuATMOSPHERESEARTHLIKEPLANETS2014}
Lupu R.~E.,  et~al., 2014, \mn@doi [The Astrophysical Journal]
  {10.1088/0004-637X/784/1/27}, 784, 27

\bibitem[\protect\citeauthoryear{{Lustig-Yaeger}, Meadows, Crisp, Line  \&
  Robinson}{{Lustig-Yaeger}
  et~al.}{2023}]{lustig-yaegerEarthTransitingExoplanet2023}
{Lustig-Yaeger} J.,  Meadows V.~S.,  Crisp D.,  Line M.~R.,   Robinson T.~D.,
  2023, \mn@doi [The Planetary Science Journal] {10.3847/PSJ/acf3e5}, 4, 170

\bibitem[\protect\citeauthoryear{MacDonald}{MacDonald}{2023}]{macdonaldPOSEIDONMultidimensionalAtmospheric2023}
MacDonald R.~J.,  2023, \mn@doi [Journal of Open Source Software]
  {10.21105/joss.04873}, 8, 4873

\bibitem[\protect\citeauthoryear{MacDonald \& Madhusudhan}{MacDonald \&
  Madhusudhan}{2017}]{macdonaldHD209458bNew2017}
MacDonald R.~J.,  Madhusudhan N.,  2017, \mn@doi [Monthly Notices of the Royal
  Astronomical Society] {10.1093/mnras/stx804}, 469, 1979

\bibitem[\protect\citeauthoryear{Madhusudhan, Sarkar, Constantinou, Holmberg,
  Piette  \& Moses}{Madhusudhan
  et~al.}{2023}]{madhusudhanCarbonbearingMoleculesPossible2023}
Madhusudhan N.,  Sarkar S.,  Constantinou S.,  Holmberg M.,  Piette A. A.~A.,
  Moses J.~I.,  2023, \mn@doi [The Astrophysical Journal Letters]
  {10.3847/2041-8213/acf577}, 956, L13

\bibitem[\protect\citeauthoryear{{Marquez-Neila}, Fisher, Sznitman  \&
  Heng}{{Marquez-Neila}
  et~al.}{2018}]{marquez-neilaSupervisedMachineLearning2018}
{Marquez-Neila} P.,  Fisher C.,  Sznitman R.,   Heng K.,  2018, Supervised
  {{Machine Learning}} for {{Analysing Spectra}} of {{Exoplanetary
  Atmospheres}} (\mn@eprint {arXiv} {1806.03944})

\bibitem[\protect\citeauthoryear{Matchev, Matcheva  \& Roman}{Matchev
  et~al.}{2022a}]{matchevUnsupervisedMachineLearning2022}
Matchev K.~T.,  Matcheva K.,   Roman A.,  2022a, \mn@doi [The Planetary Science
  Journal] {10.3847/PSJ/ac880b}, 3, 205

\bibitem[\protect\citeauthoryear{Matchev, Matcheva  \& Roman}{Matchev
  et~al.}{2022b}]{matchevAnalyticalModelingExoplanet2022}
Matchev K.~T.,  Matcheva K.,   Roman A.,  2022b, \mn@doi [The Astrophysical
  Journal] {10.3847/1538-4357/ac610c}, 930, 33

\bibitem[\protect\citeauthoryear{Meadows et~al.,}{Meadows
  et~al.}{2018}]{meadowsExoplanetBiosignaturesUnderstanding2018}
Meadows V.~S.,  et~al., 2018, \mn@doi [Astrobiology] {10.1089/ast.2017.1727},
  18, 630

\bibitem[\protect\citeauthoryear{Munsaket, Awiphan, Chainakun  \&
  Kerins}{Munsaket et~al.}{2021}]{munsaketRetrievingExoplanetAtmospheric2021}
Munsaket P.,  Awiphan S.,  Chainakun P.,   Kerins E.,  2021, \mn@doi [Journal
  of Physics: Conference Series] {10.1088/1742-6596/2145/1/012010}, 2145,
  012010

\bibitem[\protect\citeauthoryear{Nixon \& Madhusudhan}{Nixon \&
  Madhusudhan}{2020}]{nixonAssessmentSupervisedMachine2020}
Nixon M.~C.,  Madhusudhan N.,  2020, \mn@doi [Monthly Notices of the Royal
  Astronomical Society] {10.1093/mnras/staa1150}, 496, 269

\bibitem[\protect\citeauthoryear{Pedregosa et~al.,}{Pedregosa
  et~al.}{2011}]{pedregosaScikitlearnMachineLearning2011}
Pedregosa F.,  et~al., 2011, Journal of Machine Learning Research, 12, 2825

\bibitem[\protect\citeauthoryear{Rackham \& De~Wit}{Rackham \&
  De~Wit}{2024}]{rackhamRobustCorrectionsStellar2024}
Rackham B.~V.,  De~Wit J.,  2024, \mn@doi [The Astronomical Journal]
  {10.3847/1538-3881/ad5833}, 168, 82

\bibitem[\protect\citeauthoryear{Rackham, Apai  \& Giampapa}{Rackham
  et~al.}{2018}]{rackhamTransitLightSource2018}
Rackham B.~V.,  Apai D.,   Giampapa M.~S.,  2018, \mn@doi [The Astrophysical
  Journal] {10.3847/1538-4357/aaa08c}, 853, 122

\bibitem[\protect\citeauthoryear{Rackham, Apai  \& Giampapa}{Rackham
  et~al.}{2019}]{rackhamTransitLightSource2019}
Rackham B.~V.,  Apai D.,   Giampapa M.~S.,  2019, \mn@doi [The Astronomical
  Journal] {10.3847/1538-3881/aaf892}, 157, 96

\bibitem[\protect\citeauthoryear{Rackham et~al.,}{Rackham
  et~al.}{2023}]{rackhamEffectStellarContamination2023}
Rackham B.~V.,  et~al., 2023, \mn@doi [RAS Techniques and Instruments]
  {10.1093/rasti/rzad009}, 2, 148

\bibitem[\protect\citeauthoryear{Sagan, Thompson, Carlson, Gurnett  \&
  Hord}{Sagan et~al.}{1993}]{saganSearchLifeEarth1993}
Sagan C.,  Thompson W.~R.,  Carlson R.,  Gurnett D.,   Hord C.,  1993, \mn@doi
  [Nature] {10.1038/365715a0}, 365, 715

\bibitem[\protect\citeauthoryear{Schwieterman}{Schwieterman}{2016}]{schwietermanExploringHabitabilityMarkers2016}
Schwieterman E.~W.,  2016, Thesis

\bibitem[\protect\citeauthoryear{Schwieterman \& Leung}{Schwieterman \&
  Leung}{2024}]{schwietermanOverviewExoplanetBiosignatures2024}
Schwieterman E.~W.,  Leung M.,  2024

\bibitem[\protect\citeauthoryear{Schwieterman et~al.,}{Schwieterman
  et~al.}{2018}]{schwietermanExoplanetBiosignaturesReview2018}
Schwieterman E.~W.,  et~al., 2018, \mn@doi [Astrobiology]
  {10.1089/ast.2017.1729}, 18, 663

\bibitem[\protect\citeauthoryear{Soboczenski et~al.,}{Soboczenski
  et~al.}{2018}]{soboczenskiBayesianDeepLearning2018}
Soboczenski F.,  et~al., 2018, Bayesian {{Deep Learning}} for {{Exoplanet
  Atmospheric Retrieval}} (\mn@eprint {arXiv} {1811.03390})

\bibitem[\protect\citeauthoryear{Sorower}{Sorower}{2010}]{sorowerLiteratureSurveyAlgorithms2010}
Sorower M.~S.,  2010

\bibitem[\protect\citeauthoryear{Turbet, Bolmont, Bourrier, Demory, Leconte,
  Owen  \& Wolf}{Turbet et~al.}{2020}]{turbetReviewPossiblePlanetary2020}
Turbet M.,  Bolmont E.,  Bourrier V.,  Demory B.-O.,  Leconte J.,  Owen J.,
  Wolf E.~T.,  2020, \mn@doi [Space Science Reviews]
  {10.1007/s11214-020-00719-1}, 216, 100

\bibitem[\protect\citeauthoryear{Vasist, Rozet, Absil, Molli{\`e}re, Nasedkin
  \& Louppe}{Vasist et~al.}{2023}]{vasistNeuralPosteriorEstimation2023}
Vasist M.,  Rozet F.,  Absil O.,  Molli{\`e}re P.,  Nasedkin E.,   Louppe G.,
  2023, \mn@doi [Astronomy \& Astrophysics] {10.1051/0004-6361/202245263}, 672,
  A147

\bibitem[\protect\citeauthoryear{Waldmann}{Waldmann}{2016}]{waldmannDREAMINGATMOSPHERES2016}
Waldmann I.~P.,  2016, \mn@doi [The Astrophysical Journal]
  {10.3847/0004-637X/820/2/107}, 820, 107

\bibitem[\protect\citeauthoryear{Wheatley, Louden, Bourrier, Ehrenreich  \&
  Gillon}{Wheatley et~al.}{2017}]{wheatleyStrongXUVIrradiation2017}
Wheatley P.~J.,  Louden T.,  Bourrier V.,  Ehrenreich D.,   Gillon M.,  2017,
  \mn@doi [Monthly Notices of the Royal Astronomical Society: Letters]
  {10.1093/mnrasl/slw192}, 465, L74

\bibitem[\protect\citeauthoryear{Wordsworth \& Pierrehumbert}{Wordsworth \&
  Pierrehumbert}{2014}]{wordsworthABIOTICOXYGENDOMINATEDATMOSPHERES2014}
Wordsworth R.,  Pierrehumbert R.,  2014, The Astrophysical Journal Letters

\bibitem[\protect\citeauthoryear{Wunderlich et~al.,}{Wunderlich
  et~al.}{2019}]{wunderlichDetectabilityAtmosphericFeatures2019}
Wunderlich F.,  et~al., 2019, \mn@doi [Astronomy \& Astrophysics]
  {10.1051/0004-6361/201834504}, 624, A49

\bibitem[\protect\citeauthoryear{Yip, Changeat, Nikolaou, Morvan, Edwards,
  Waldmann  \& Tinetti}{Yip et~al.}{2021}]{yipPeekingBlackBox2021}
Yip K.~H.,  Changeat Q.,  Nikolaou N.,  Morvan M.,  Edwards B.,  Waldmann
  I.~P.,   Tinetti G.,  2021, \mn@doi [The Astronomical Journal]
  {10.3847/1538-3881/ac1744}, 162, 195

\bibitem[\protect\citeauthoryear{Zingales \& Waldmann}{Zingales \&
  Waldmann}{2018}]{zingalesExoGANRetrievingExoplanetary2018}
Zingales T.,  Waldmann I.~P.,  2018, \mn@doi [The Astronomical Journal]
  {10.3847/1538-3881/aae77c}, 156, 268

\makeatother
\end{thebibliography}

\bsp	
\label{lastpage}
\end{document}